\documentclass{aa}

\usepackage{graphicx}
\usepackage{epsfig}
\usepackage{natbib}
\usepackage[varg]{txfonts}
\usepackage{hyperref}
\newcommand{\pun}[1]{\mbox{\rm\,#1}} 

\newcommand{\mlp}{\ensuremath{\alpha_{\mathrm{MLT}}}}

\newcommand{\beq}{\begin{equation}}
\newcommand{\eeq}{\end{equation}}

\newcommand{\xtmean}[1]{\ensuremath{\left\langle #1\right\rangle}}

\newcommand{\OoFe}{\ensuremath{\left[\mathrm{O}/\mathrm{Fe}\right]}}
\newcommand{\SH}{\ensuremath{S_{\mathrm{\!H}}}}
\newcommand{\logAO}{\ensuremath{{\rm A(O)}}}
\newcommand{\chisq}{\ensuremath{{\rm \chi^{2}}}}
\newcommand{\vmic}{\ensuremath{{\xi_{\mathrm{mic}}}}}

\newcommand{\COBOLD}{{\tt CO5BOLD}}
\newcommand{\LHD}{{\tt LHD}}

\newcommand{\ATLAS}{{\tt ATLAS}}

\newcommand{\MARCS}{{\tt MARCS}}

\newcommand{\LINFOR}{{\tt Linfor3D}}

\newcommand{\NLTETD}{{\tt NLTE3D}}
\newcommand{\IONOPA}{{\tt IONOPA}}

\begin{document}

   \title{The photospheric solar oxygen project:}

   \subtitle{IV. 3D-NLTE investigation of the 777\,nm triplet lines}

   \author{M.\,Steffen\inst{1,3}
   \and D.\,Prakapavi\v{c}ius\inst{2}
   \and E.\,Caffau\inst{3}
   \and H.-G.\,Ludwig\inst{4,3}
   \and P.\,Bonifacio\inst{3}
   \and R.\,Cayrel\inst{3}
   \and A.\,Ku\v{c}inskas\inst{2,5}
   \and W.C.\,Livingston\inst{6}
   }

   \institute{Leibniz-Institut f\"ur Astrophysik Potsdam, 
              An der Sternwarte 16, D-14482 Potsdam, Germany,
              \email{msteffen@aip.de}
   \and Institute of Theoretical Physics and Astronomy, Vilnius University, 
        A. Go\v {s}tauto 12, Vilnius LT-01108, Lithuania
   \and GEPI, Observatoire de Paris, CNRS, Universit\'{e} Paris Diderot, 
        Place Jules Janssen, 92190 Meudon, France
   \and ZAH Landessternwarte K\"{o}nigstuhl, D-69117 Heidelberg, Germany
   \and Astronomical Observatory, Vilnius University, M. K. \v{C}iurlionio 29, 
   Vilnius LT-03100, Lithuania
   \and National Solar Observatory, 950 North Cherry Avenue, Tucson, 
        AZ 85719, USA
   }
   \date{Received XXX; accepted XXX}

  \abstract
   {The solar photospheric oxygen abundance is still widely debated. Adopting
    the solar chemical composition based on the `low' oxygen abundance, as 
    determined with the use of three-dimensional (3D) hydrodynamical model 
    atmospheres, results in a well-known mismatch between theoretical solar 
    models and helioseismic measurements that is so far unresolved.}
   {We carry out an independent redetermination of the solar oxygen abundance 
    by investigating the center-to-limb variation of the \ion{O}{i} IR 
    triplet lines at 777\,nm in different sets of spectra.}
   {The high-resolution and high signal-to-noise solar center-to-limb spectra 
   are analyzed with the help of detailed synthetic line profiles based on 3D
   hydrodynamical \COBOLD\ model atmospheres and 3D non-LTE line formation 
   calculations with \NLTETD. The idea is to exploit the information
   contained in the observations at different limb angles to simultaneously
   derive the oxygen abundance, \logAO, and the scaling factor \SH\ that
   describes the cross-sections for inelastic collisions with neutral 
   hydrogen relative the classical Drawin formula. Using the same 
   codes and methods, we compare our 3D results with those obtained from the 
   semi-empirical Holweger-M\"uller model atmosphere as well as from different 
   one-dimensional (1D) reference models.}
   {With the \COBOLD\ 3D solar model, the best fit of the center-to-limb 
    variation of the triplet lines is obtained when the collisions 
    by neutral hydrogen atoms are assumed to be efficient, i.e.,\ when the 
    scaling factor \SH\ is between 1.2 and 1.8, depending on the choice of 
    the observed spectrum and the triplet component used in the analysis. 
    The line profile fits achieved with standard 1D model atmospheres (with 
    fixed microturbulence, independent of disk position $\mu$) are  clearly of 
    inferior quality compared to the 3D case, and give the best match 
    to the observations when ignoring collisions with neutral hydrogen 
    (\SH=0). The results derived with the Holweger-M\"uller model are 
    intermediate between 3D and standard 1D.}
   {The analysis of  various observations of the triplet lines with 
    different methods yields oxygen abundance values (on a logarithmic scale
    where A(H)=12) that fall in the range $8.74 < \logAO < 8.78$, 
    and our best estimate of the 3D
    non-LTE solar oxygen abundance is $\logAO = 8.76\pm0.02$.
    All 1D non-LTE models give much lower oxygen abundances, by up to 
    $-0.15$\,dex. This is mainly a consequence of the assumption of a 
    $\mu$-independent microturbulence. An independent determination of
    the relevant collisional cross-sections is essential to substantially 
    improve the accuracy of the oxygen abundance derived from the 
    \ion{O}{i} IR triplet.}


\authorrunning{M. Steffen et al.}
\titlerunning{Solar oxygen abundance determined from the 777 triplet lines}

  \maketitle

\section{Introduction}
  Oxygen, the most abundant chemical element of stellar nucleosynthetic
  origin, is a widely used tracer for the evolution of various stellar
  populations. Oxygen is mainly produced in massive stars that end their
  evolution as Type II supernovae. Since the main production of 
  galactic iron by type Ia supernovae from less massive progenitor stars
  only contributes  after some delay, the oxygen-to-iron abundance ratio, 
  \OoFe\footnote{\OoFe\ = 
  $\log[$N(O)/N(Fe)$]$\(_\star\)$-\log[$N(O)/N(Fe)$]$\(_\odot\).}, 
  can be used as an 
  indicator of the chemical evolution of a particular composite stellar 
  population. This abundance ratio, along with other $\alpha$-element 
  abundances, has been used to investigate Galactic star formation rates 
  and to constrain models of the Galactic Chemical Evolution 
  \citep{MF92, Gra00}. The solar oxygen abundance, 
  \logAO\footnote{\logAO\ = $\log[$N(O)/N(H)$]+12$}, 
  is a natural reference for these investigations.
  
  Since oxygen is a volatile element, it condenses only partly in meteoritic 
  matter, so that the relative amount of oxygen present in meteorites
  (e.g., with respect to a reference element like silicon) is not
  representative of the oxygen abundance ratio that was actually
  present in the original solar nebula. But it is thought that the latter 
  is preserved in the solar photosphere, and hence can be inferred by
  spectroscopic analysis. Recent state-of-the-art solar spectroscopic
  abundance determinations resulted in a rather low
  oxygen abundance of $\logAO = 8.66-8.68$\,dex \citep{PAK09}. Previous
  estimates by \citet{AGS04}, who analyzed atomic and molecular oxygen 
  lines in disk-integrated spectra, yielded a consistent value,
  $\logAO = 8.66 \pm 0.05$\,dex. The work of \citet{ALA01}, who analyzed the 
  forbidden oxygen line at 630\,nm, resulted in $\logAO = 8.69 \pm 0.05$\,dex, 
  while the investigation of \citet{AAF04}, who performed a \chisq\ analysis 
  of the 3D non-LTE center-to-limb variation of the O~I IR triplet, suggested 
  a slightly higher value of $\logAO = 8.72$\,dex. 

  The rather low value of the solar oxygen abundance, $\logAO \la 8.70$\,dex,
  as found in the above mentioned investigations, poses significant problems for
  theoretical solar structure models to meet the helioseismic constraints. 
  These models depend on the solar chemical composition via radiative 
  opacities, equation of state, and energy production rates. Most notably, 
  changes in CNO abundances influence the structure of the radiative interior,
  the location of the base of convective envelope, the helium abundance in the 
  convection zone, and the sound speed profile. It was shown by \citet{BA08} 
  that the `old' solar chemical composition of \citet{GS98}, where 
  $\logAO = 8.83 \pm 0.06$, yields a much better agreement between 
  theory and helioseismic observations than the new solar metallicity
  with a significantly reduced oxygen abundance of $\logAO \la 8.70$.

  According to our own previous studies 
  \citep[][hereafter Paper~I]{CLS08}, the solar oxygen abundance is
  somewhat higher, $\logAO = 8.73-8.79$\,dex, with a weighed mean of 
  $\logAO = 8.76 \pm 0.07$\,dex. This rather high oxygen abundance was 
  supported by the work of \citet{ALL13}, who used a
  \COBOLD\ model atmosphere to analyze molecular CO lines 
  (assuming $n$(C)/$n$(O)=0.5) and 
  derived $\logAO = 8.78 \pm 0.02$\,dex. \citet{VSD14} has shown that 
  the chemical composition presented by \citet{CLS11}, which includes a
  `high' \logAO, is in much better agreement with the overall metal abundance
  inferred from helioseismology and solar neutrino data.
  
  Spectroscopic abundance analysis relies on stellar atmosphere models 
  and spectrum synthesis, as a theoretical basis. The reliability of these 
  abundance determinations depends on the physical realism of the atmosphere 
  models and of the spectral line formation theory. On the observational side, 
  high signal-to-noise spectra of sufficient spectral resolution are
  essential, as is the availability of clean spectral lines. The number of
  suitable oxygen lines in the optical solar spectrum is very limited.
  The \ion{O}{i} IR triplet lines at 777\,nm are essentially free from any 
  contamination, and hence could be expected to be a good abundance indicator.
  It is well known, however, that the triplet lines are prone to departures 
  from local thermodynamic equilibrium (LTE) in solar-type stars
  \citep[e.g.,][]{Kis91}.
  The non-LTE effects lead to stronger synthetic spectral lines than in LTE, 
  and hence a lower oxygen abundance is required to match the observed line
  strength. 

  It has been repeatedly demonstrated that it is impossible to reproduce the
  center-to-limb variation of the triplet lines under the assumption of LTE
  \citep[][]{Kis93, PAK09}. \citet{PAK09} have also shown that non-LTE
  synthetic line profiles based on standard one-dimensional (1D) model 
  atmospheres (e.g., \ATLAS, \MARCS) cannot explain the observed properties 
  of the triplet, while the semi-empirical Holweger-M\"uller model
  \citep{Hol67, HM74} is able to achieve a consistent match of the
  observations. Three-dimensional (3D) model atmospheres, however, 
  give the best results.

  In this paper, we expand on the work done in Paper~I, following up on
  the investigation of the solar \ion{O}{i} IR triplet at 777\,nm with an 
  independent 3D,~non-LTE analysis of several high-quality 
  solar intensity spectra for different values of the direction cosine 
  $\mu$\footnote{$\mu = \cos\theta$, where $\theta$ is the angle between 
  the line of sight and the direction perpendicular to the solar surface; 
  $\mu = 1$ at disk center, while $\mu = 0$ at the limb.} (hereafter 
  $\mu$-angle for short) and of a spectrum of the disk-integrated flux.
  In Section~\ref{sect:solar-obs}, we describe the spectra that
  we use for our analysis, and in Sect.\,\ref{sect:theory} we outline the
  theoretical foundations of interpreting the spectra. The main results are
  presented in Sect.\,\ref{sect:Solar-abu}, in which we describe the adopted
  methodology, perform different comparisons between observed and theoretical 
  spectra, and derive the most likely value of the solar oxygen abundance. We 
  discuss the results and their implications in Sect.\,\ref{sect:discussion}, 
  before concluding the paper with a summary of our findings in 
  Sect.\,\ref{sect:conclusions}.

\section{Observations \label{sect:solar-obs}}

  We utilize intensity spectra of the \ion{O}{i} IR triplet
  recorded at different $\mu$-angles across the solar disk. As 
  mentioned above, the analysis of the triplet lines requires a detailed
  modeling of their departures from LTE, which is rather sensitive to the 
  strength of collisions with neutral hydrogen atoms. Since the relevant 
  collisional cross-sections are essentially unknown, we derive 
  both the oxygen abundance, \logAO, and the scaling factor with respect to 
  the classical (Drawin) cross-sections for collisional excitation and 
  ionization of oxygen atoms by neutral hydrogen, \SH,
  from a simultaneous fit of 
  the line profiles or their equivalent widths at different $\mu$-angles.
  An analysis of the disk-center or disk-integrated spectrum alone would 
  be insufficient to yield unique results. Since the impact of \SH\ depends
  systematically on the depth of line formation and therefore on
  $\mu$, the variation of the spectral lines with disk position adds another 
  dimension to the analysis and helps to constrain the problem. 
  Following a similar idea, \citet{caffau15} utilized the
  center-to-limb variation of the oxygen and nickel blend at 630\,nm to 
  separate the contribution of the individual elements.
   
  We make use of the following three independent 
  observational data sets: 

\begin{enumerate}

\item
The absolutely calibrated FTS spectra obtained at Kitt Peak in the 1980s, 
as described by \citet{neckelobs} and \citet{neckel1999},
covering the wavelength range from 330\pun{nm} to 1250\pun{nm} for 
the disk center and full disk (Sun-as-a-star), respectively. In the following, 
we refer to this as ``Neckel Intensity'' and ``Neckel Flux''. The  spectral 
purity ($\Delta\lambda$) ranges from 0.4\pun{pm} at 330\pun{nm} to 2\pun{pm} at 
1250\pun{nm} (825\,000 $>$ $\lambda/\Delta\lambda$ $>$ 625\,000).\\

\item
A set of spectra observed by one of the authors (WCL) at Kitt Peak with the
McMath-Pierce Solar Telescope and rapid scan double-pass spectrometer
\citep[see][and references therein]{Livingston2007}. Observations were taken 
in single-pass mode on 12 September, 2006; the local time was about 10h15m 
to 10h45m. The resolution was set by the slit width: 0.1\,mm for the entrance
and 0.2\,mm for the exit slit. This yields a resolution of 
$\lambda/\Delta\lambda \approx 90\,000$. The image scale is
$2$\,\farcs$5$\,mm$^{-1}$. The slit has a length of 10\,mm and was set parallel
to the limb. Observations were made at six $\mu$-angles: $\mu = 1.00,~ 0.87,~
0.66,~ 0.48,~ 0.35,~ {\rm and}~ 0.25$, along the meridian N to S in 
telescope (geographic) coordinates. We have acquired two spectra for any
$\mu$ value and four spectra at  disk center. The exposure time for
each individual spectrum is 100\,s, thus essentially freezing the five-minute
solar oscillations. No wavelength calibration source was observed, and we used
the available telluric lines to perform the wavelength calibration. The 
positions of the telluric lines were measured on the Neckel Intensity 
spectrum and a linear dispersion relation was assumed.

After the wavelength calibration, we summed the spectra at equal $\mu$ to 
improve the signal-to-noise ratio (S/N). The continuum of the summed spectra 
was defined by a cubic spline passing through a predefined set of line-free 
spectral windows. A comparison of the normalized 
disk-center spectrum with the Neckel Intensity atlas reveals a systematic
difference in the equivalent widths, in the sense that the spectral lines
are weaker in our own spectrum by about 4\%. We attribute this difference to 
instrumental stray light, and correct the normalized spectra for all 
$\mu$-angles as $I_{\rm corr}(\lambda) = (1+f)\,I(\lambda) - f$, where
$f=0.0405$ is the stray light fraction. We estimate the signal-to-noise
ratio of the stray light corrected spectra to be S/N $\approx 1500$ at
disk center, and S/N $\approx 700$ at $\mu=0.25$. In the following, we refer 
to these spectra as WCLC.\\

\item
The high-quality spatially and temporally averaged intensity
spectra used by \citet{PAK09}\footnote{\url{http://cdsarc.u-strasbg.fr/
viz-bin/qcat?J/A+A/508/1403}}, recorded with the Swedish 1-m Solar Telescope 
\citep{SBK03} on Roque de Los Muchachos at La Palma over two weeks in May
2007. For these observations, the TRIPPEL spectrograph was used with a slit 
width set to 25\,${\rm\mu}$m, corresponding to 0.11\arcsec\ on the sky. 
The spectral resolution of this data set is $\lambda/\Delta\lambda \approx 
200\,000$. The data set consists of spectra at five disk positions,  
$\mu = 1.00$, $0.816$, $0.608$, $0.424$, and $0.197$. Since a large number 
of spectra at different instances of time and space were summed up, a high 
signal-to-noise ratio of S/N $\approx 1\,200-1\,500$ was achieved at all 
$\mu$-angles. The spectra are corrected for stray light.
See \citet{PKA09,PAK09} for a 
detailed discussion regarding this observational data set.

\end{enumerate}

\section{Stellar atmospheres and spectrum synthesis\label{sect:theory}}
To compare observed and theoretical spectra, we performed
detailed non-LTE spectral line profile calculations. In this section, we
describe the theoretical basis of the calculations performed for the present
analysis of the center-to-limb variation of the different solar spectra.

\subsection{Model atmospheres}
We give an overview of the properties of the 
3D hydrodynamic and 1D hydrostatic model 
atmospheres used in this paper.

\subsubsection{3D hydrodynamical \COBOLD\ model atmospheres
\label{sect:3D_models}}
The 3D hydrodynamical solar model atmosphere utilized in this work
is very similar to that used in Paper~I\footnote{The only difference
is that we now employ an equation of state that is 
consistent with the chemical abundance mix of the opacity tables.}. 
It was computed with the \COBOLD\ code \citet{FSL12} on a Cartesian grid 
consisting of $140\times 140$ grid points horizontally and $150$ grid 
points vertically, spanning $5.6\times 5.6$~Mm$^2$ horizontally and 
$2.25$~Mm vertically (``box-in-a-star'' setup). Open boundary conditions 
were applied vertically, allowing matter and radiation to flow in and out 
of the computational domain while conserving the total mass of matter 
inside the box. At the lower boundary, we prescribe the entropy of the 
matter entering the model from below, which controls the outgoing 
bolometric radiative flux (effective temperature) of the model atmosphere. 
At the top boundary, the temperature of inflowing matter and the scale heights 
of pressure and optical depth beyond the upper boundary are controlled 
via free parameters. In the lateral directions, periodic boundary conditions 
are applied. Under these conditions, matter and radiation that exit the
model nonvertically on one side, enter the box from the other side.
   
\begin{figure}[tb]
\centering
\mbox{\includegraphics[bb=56 0 720 600,width=9cm]{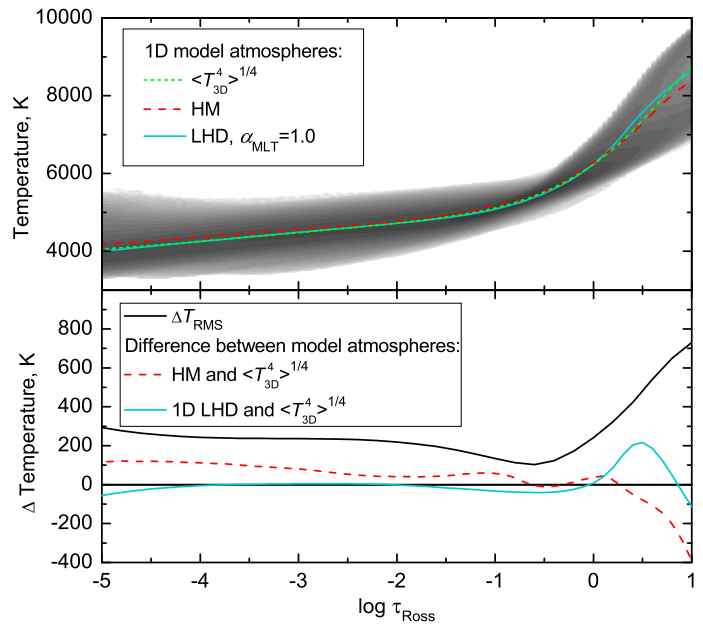}}
\caption{\emph{Top}: temperature structure of 3D, \xtmean{3D}, 1D \LHD, and 
HM model atmosphere on the Rosseland optical depth scale. \emph{Bottom}: 
horizontal temperature fluctuations, $\Delta T_{\rm RMS}$, of the 3D model
at constant $\tau_{\rm Ross}$ (black) and temperature difference 
\mbox{HM - \xtmean{3D}} (dashed) and \mbox{1D \LHD - \xtmean{3D}} (blue) 
as a function of $\log \tau_{\rm Ross}$.}
\label{fig:Ttau}
\end{figure}

The construction of the opacity tables that are needed for the radiation 
transport relies on monochromatic \MARCS\ opacities \citep{GEK08}, computed 
in LTE and adopting the solar chemical composition of \citet{GS98}, with the 
exception of the CNO abundances that were set to those of \citet{AGS05}. 
To speed up the calculations, opacities are grouped into 12 opacity bins according to the ideas of \citet{NORD82}, and further elaboration
by \citet{LJS94} and \citet{VBS04}. In the calculation of the tables, the 
continuum scattering opacity was added to the true absorption opacity.
      
This setup was evolved in time until a thermally relaxed, statistically
steady state was reached. Subsequently, the run was continued for another
two hours of stellar time, covering about 24 convective turnover times.
Out of this two-hour time sequence, 20 representative model snapshots were
chosen so that their mean effective temperature (radiative flux)
and its temporal RMS fluctuation closely agree with the corresponding values 
of the fully resolved time sequence. In addition, we made sure that the 
horizontally averaged vertical velocity showed a similar depth dependence 
and temporal fluctuation in the subsample as in the whole sample. It was 
argued previously \citep{caffau09,ludwig09} that such subsample of model 
atmospheres represents the whole evolution in terms of spectral line 
properties reasonably well. By using the
selected snapshots, we can thus considerably reduce the computation cost of 
the subsequent spectrum synthesis. For additional information about the 
3D solar model used in this paper and comparison with other 
3D-hydrodynamical solar model atmospheres, see~\citet{BCS12}. 

We obtain the so-called \xtmean{3D} model
via horizontal and temporal averaging of temperature ($T^4$) and gas pressure 
($P$) on surfaces of constant optical depth ($\tau_{\rm Ross}$).  
The \xtmean{3D} model is a 1D atmosphere representing the mean
properties of the 3D model (cf.\ Fig.\,\ref{fig:Ttau}). Comparison of the 
line formation properties of the full 3D and the \xtmean{3D} model gives 
an indication of the importance of horizontal inhomogeneities in the 
process of line formation. In the present work, the performance of 
the \xtmean{3D} model is compared with that of the standard 1D models.

\subsubsection{1D \LHD\ model atmospheres\label{sect:1D_models}}

In the present analysis, we have also used classical 1D model atmospheres
computed with the \LHD\ code \citep[see][]{CLS08}. 
We computed ID  \LHD\ and 3D \COBOLD\ model atmospheres  using identical 
stellar parameters, chemical composition, equation of state, opacity binning 
schemes, and radiation transport methods, and thus these models are suitable 
for the analysis of 3D--1D effects through a differential comparison. 
Since \LHD\ model atmospheres are static, convection is treated via the
mixing length theory \citep[MLT,][]{BV58, mihalas78}.
   
Since the oxygen triplet lines have a high excitation potential and, hence, 
form in the deep photosphere, it may be expected that their strength is 
sensitive to the choice of mixing length parameter, \mlp. To probe 
the possible influence of \mlp, we have used a grid of \LHD\ model atmospheres 
with $\mlp = 0.5, 1.0, 1.5$. After performing some tests, however, we found
negligible influence of \mlp\ on the triplet lines 
(see Appendix~\ref{app:mlp} for details). Hence, all results shown in the
following are based in the \LHD\ model atmosphere with $\mlp = 1.0$.
   
\subsubsection{Semi-empirical Holweger-M\"uller solar model atmosphere}
\label{sect:HMull}

We have also used the semi-empirical model atmosphere derived by 
\citet{HM74} \citep[in the following HM model, see also][]{Hol67}. 
The $T(\tau)$ relation 
of this model was empirically fine-tuned to match observations of the 
solar continuum radiation and a variety of spectral lines.

The temperature structure of the HM model is significantly different
from that of the theoretical 1D \LHD\ model atmospheres, as illustrated
in Fig.\,\ref{fig:Ttau}. In the optical depth range  
$-5.0 < \log \tau_{\rm Ros} < -0.5$, 
the HM model is slightly warmer than the \xtmean{3D} model by $50$ to 
$100$~K, while the 1D \LHD\ model almost perfectly matches the 
\xtmean{3D} model. In the deep photosphere 
($-0.5 < \log \tau_{\rm Ros} < +0.5$), where significant parts of the triplet 
lines originate, the HM model closely follows the temperature of the 
\xtmean{3D} model, while the 1D \LHD\ shows a somewhat steeper
temperature gradient in these layers. This results in stronger OI triplet
lines computed from the \LHD\ model compared to those of the HM atmosphere.

\subsection{3D/1D non-LTE spectrum synthesis}
\label{sect:synth}

Theoretical spectral lines were synthesized in two separate steps. Firstly,
the departure coefficients\footnote{The departure coefficient of
atomic level $i$, $b_{i}$, is defined as the ratio of non-LTE to LTE
population number density, $n_{i}/n_{i}^\ast$} of the atomic levels involved in 
the \ion{O}{i} triplet were computed with the \NLTETD\ code. Then the 
departure coefficients
were passed to the spectral line synthesis package \LINFOR\ that was used to compute spectral line profiles across the solar disk. In this 
section, we describe the basic assumptions and methods applied in the non-LTE 
spectrum synthesis calculations.

\subsubsection{The oxygen model atom} 
\label{sect:modelatom}

\begin{figure}[!h]
\centering
\includegraphics[width=9cm]{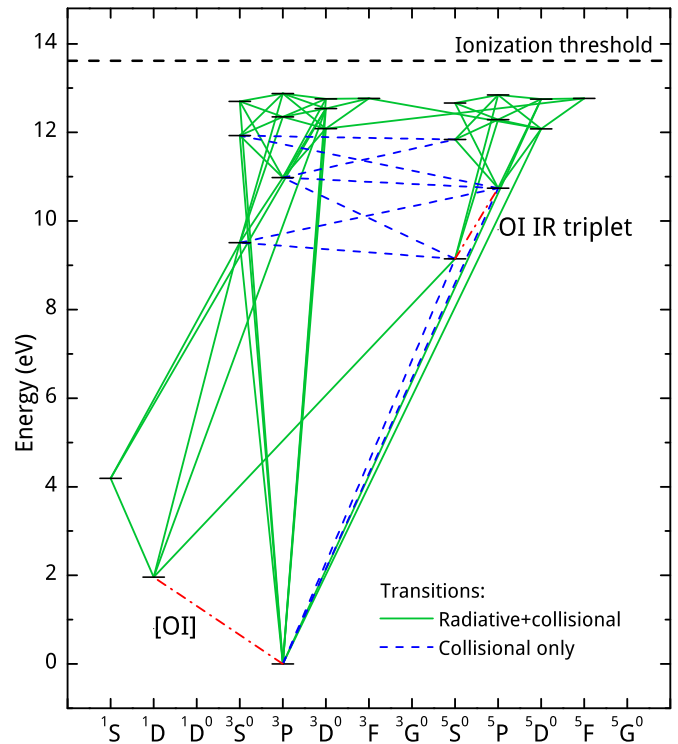}
\caption{Oxygen model atom used in this work. Radiatively allowed transitions 
         are marked with green solid  lines. Radiatively forbidden 
         transitions, which were treated via electron-impact excitation only,
         are marked by  blue dashed lines. The forbidden [\ion{O}{i}] line 
         and the \ion{O}{i} IR triplet are marked by red dash-dotted lines. 
         We took collisional and radiative bound-free transitions into 
         account for each level.}
\label{fig:Grotrtian_OI}
\end{figure}

We use an oxygen model atom that consists of 22 levels of 
\ion{O}{i} (plus continuum), each having an associated bound-free transition.
In addition, we take 64 bound-bound transitions into account, some of which are
purely collisional. Fig.~\ref{fig:Grotrtian_OI} shows the Grotrian diagram of 
our model atom.
  
Level energies and statistical weights of the model atom were taken from the 
NIST Atomic Spectra Database 
Version~5\footnote{\url{http://www.nist.gov/pml/data/asd.cfm/}} 
(see Tab.\,\ref{tab:oxygen-levels}).
Cross sections for radiative, bound-free transitions were taken from the
TOPBASE service\footnote{\url{http://cdsweb.u-strasbg.fr/topbase/topbase.html}} 
of the Opacity Project \citep{CMO93}. Lacking reliable collisional 
cross sections derived from laboratory measurements or quantum-mechanical 
calculations, we resort to the classical prescriptions of
Allen \citep[][p. 41]{allen73} and Drawin \citep{DRAW69, SH84, Lam93}
to obtain the rate coefficients for bound-free transitions 
due to ionizing collisions with electrons and neutral hydrogen, respectively. 
As the Drawin approximation is highly uncertain (see 
Sect.\,\ref{sect:discussion3} below), 
we scale the rates obtained from the Drawin formula by the adjustable
parameter \SH\ , which is deduced simultaneously 
with the oxygen abundance (see Sect.\,\ref{sect:Solar-abu}). 
Finally, we include the charge transfer reaction
$\mathrm{O^{0}~+~H^{+}~\rightleftharpoons~O^{+}~+~H^{0}}$ for the ground level
of oxygen, as described by \citet{AR85}. Because the ionization potentials
of \ion{O}{i} and H~I are similar, this reaction is nearly resonant and very efficient.
It ensures that the ground level of \ion{O}{i} is in equilibrium with
\ion{O}{ii}.

Out of the 64 bound-bound transitions, 54 are radiatively and collisionally
permitted, while the remaining ten are radiatively forbidden. For the
latter transitions, we only take collisions with electrons  into account 
(see below). Einstein coefficients and wavelengths for the permitted 
transitions were again taken from the NIST Atomic Spectra Database
(see Tab.\,\ref{tab:oxygen-trans}).

Rate coefficients for collisional excitation by neutral hydrogen were computed
with classical Drawin formula and scaled by the same factor \SH\ as applied
to the ionizing collisions by neutral hydrogen. Since the collisional cross 
section according to the Drawin formalism is proportional to the oscillator 
strength $gf$ of the transition, the levels of the radiatively forbidden 
transitions are also not coupled by hydrogen collisional excitation in our 
model atom. Alternative assumptions and their impact on the derived
oxygen abundance are briefly discussed in Sect.\,\ref{sect:discussion31} below.
 
Rate coefficients for the collisional excitation
by e$^{-}$ were taken from the detailed R-matrix computations of
\citet{Bark07}. For the radiatively allowed transitions that were not included
in the latter work, we applied the classical van Regemorter's formula
\citep{rege62}. The rate coefficients of
\citet{Bark07}, unlike those computed by the oscillator strength-dependent
classical van Regemorter's formula, are nonzero for radiatively forbidden
transitions (blue dashed lines in Fig.~\ref{fig:Grotrtian_OI}), and hence we
were able to account for the inter-system collisional transitions. 
The latter transitions are thought to play an important role in defining the
statistical equilibrium of oxygen \citep[especially at low metallicities, 
see][]{FAB09} and must be taken into account in realistic NLTE simulations.
While the work of \citet{Bark07} provides data for a total of 171 
bound-bound transitions, we have only used 48 of them, based on
the magnitude of their rate coefficient. We have performed 
tests that evaluate the influence of the remaining purely collisional 
transitions and found that it is negligible in the case of the Sun.

It should be noted that fine-structure splitting is ignored
in our model atom, and all multiplets are treated as singlets. In particular,
the upper level of the \ion{O}{i} IR triplet, level 3p$^{5}$P, is treated as 
a single superlevel with a single departure coefficient. However, since
the three triplet components are well separated in wavelength, the line 
opacity computed from the combined $gf$-value of the superlevel transition
would be too large, resulting in a wrong mean radiation field  
and, in turn, in wrong radiative excitation rates. To avoid this problem, 
we replace -- for the radiative line transfer calculations only -- 
the summed-up $gf$-value of the combined multiplet
transition with the $gf$-value of the intermediate triplet component, thus 
diminishing the line opacity by a factor 3 (see Appendix~\ref{app:finesplit} 
for additional information). Similar opacity scaling is applied for all other 
lines that involve multiplet transitions. This approach significantly 
reduces the computational cost of solving the statistical equilibrium equations.

\subsubsection{Computation of departure coefficients with \NLTETD}
\label{sect:nlte}

As a first step towards computing 3D/1D non-LTE line profiles of
the \ion{O}{i} IR triplet, we  computed departure coefficients
$b_{i}$,  using the \NLTETD\ code with the 3D/1D solar model atmospheres 
and oxygen model atom described in the previous sections.

Firstly, \NLTETD\ computes the line-blanketed radiation field that is required 
for the calculation of the photoionization transition probabilities. For this 
task, we utilized the line opacity distribution functions from \citet{CK04} 
and the continuum opacities from the \IONOPA\ package. The radiation field
is derived with a long-characteristics method where the transfer equation 
is solved with a modified Feautrier scheme. To save
computing time, continuum scattering is treated as true absorption. For solar
conditions, the validity of the latter approximation has been verified via
comparison with test calculations for a single 3D snapshot in which coherent
isotropic continuum scattering is treated consistently. To further minimize
the computational cost, we have reduced the spatial resolution
of the 3D model by using only every third point in both horizontal directions,
thus saving about 90$\%$ of the computing time.  Tests have shown that this loss
of spatial resolution does not influence the final results in any significant
way (see Appendix~\ref{app:skipping}).

Based on this fixed 3D radiation field, we compute the photoionization
and photorecombination transition probabilities. They are assumed to be 
independent of possible departures from LTE in the level population of 
\ion{O}{i}.
Then, \NLTETD\ computes the angle-averaged continuum intensities, $J_{\rm cont}$,
at the central wavelength of all of the line transitions. If a particular 
spectral line is weak, the profile averaged mean line intensity, $\bar{J}$ is 
fixed to $J_{\rm cont}$, and the radiative bound-bound transition probabilities 
of these lines are only computed once. Finally, we compute the collisional 
transition probabilities, which depend only on the local temperature.
All of those transition probabilities are fixed, as they do not depend on the 
departure coefficients. In this situation, the solution of the statistical 
equilibrium equations would be straightforward, and the departure coefficients 
of all levels could be obtained immediately, i.e.,\ without any iterative 
procedure.

However, as soon as the line transitions are no longer optically thin, the
mean radiative line intensities, $\bar{J}$, and the departure coefficients, 
$b_{i}$, are inter-dependent quantities that exhibit a highly non-linear 
coupling. In this case, we must compute $\bar{J}$ as the weighted average of 
$J_\nu$ over the local absorption profile, taking into account that line 
opacity and source function depend on the departure coefficients
of the energy levels of the transition. In the present version of
\NLTETD, the local line profile is assumed to be Gaussian, representing the 
thermal line broadening plus the effect of microturbulence. At this point,
the 3D hydrodynamical velocity field is not used, so Doppler shifts along
the line of sight are ignored. Instead, the velocity field is represented by 
a depth-independent microturbulence of $1.0$~km/s. Radiative damping, as well
as van der Waals and Stark broadening, are ignored at this stage. We verified that this simplification has a negligible influence on the resulting 
departure coefficients. For the spectrum synthesis, however, the line
broadening is treated in full detail and correctly includes the effects 
of the 3D hydrodynamical velocity field (see below).

Since it would be very
computationally expensive to use complete linearization methods
\citep[][Sect.\,8]{Auer73, Kubat10} with 3D model atmospheres, we choose to
iterate between mean radiative line intensities and departure
coefficients. However, as discussed by \citet{PSK13}, the solution of 
the oxygen non-LTE problem with \NLTETD\ by ordinary $\Lambda$-iteration
exhibits a poor convergence, which could potentially lead to faulty 
results. Hence, we implemented the accelerated $\Lambda$-iteration 
(ALI) scheme of \citet[][Sect.\,2.2]{RH91}  to achieve faster 
convergence.
  
Iterations are started with LTE line opacities and source functions,
from which the mean line intensities, $\bar{J}$, radiative transition
probabilities, and equivalent widths (EWs) of the emergent flux line profiles
are calculated. 
Then, all previously computed transition probabilities enter the equations of 
statistical equilibrium, which are solved for the unknown departure
coefficients. Using the computed departure coefficients, the non-LTE line
opacities and source functions are computed and the radiation transfer is solved
for all relevant lines using these quantities. Then the $\bar{J}$, flux EWs, 
and approximate $\Lambda$-operators ($\Lambda^{*}$) are recomputed with non-LTE
line opacities and source functions. The updated mean radiative intensities 
enter the equations of statistical equilibrium and the cycle begins again.
  
The accelerated $\Lambda$-iterations are continued until the relative changes 
(from one iteration to the next) in the flux EWs of all sufficiently strong 
lines have converged to less than $10^{-3}$. We  checked that increasing the 
accuracy of the convergence criterion does not alter the resulting
departure coefficients significantly, which might be ascribed to the good
convergence properties of the accelerated $\Lambda$-iteration scheme.

In preparation for the following investigation, we  computed departure
coefficients for a grid of \SH\ and \logAO\ values. The \SH\ values
range from 0 to 8/3 in steps of 1/3, and \logAO\ varies from 8.65 to
8.83\,dex in steps of 0.03\,dex. The spectrum synthesis calculations
described in Sect.\,\ref{sect:specsyn} are based on this rather extensive 
and dense grid of departure coefficients.

Since the \NLTETD\ code can only work with 3D model atmospheres defined on
a geometrical grid, the 1D \LHD\ and HM model atmospheres were converted 
to pseudo-3D models by replicating the particular 1D model atmosphere three 
times horizontally in each direction so that the resulting model 
would have 3x3 identical columns. Using the same equation of state and 
Rosseland opacity tables as in the 3D models, the vertical grid spacing 
of the pseudo-3D models is adjusted to ensure that the resulting model 
atmosphere is sampled at equidistant log $\tau_{\rm Ross}$ levels, and preserves 
the $T(\tau)$ relation of the original 1D model.

\subsubsection{3D/1D NLTE spectrum synthesis with \LINFOR}
\label{sect:specsyn}

The second step in computing the synthetic line profiles of the oxygen
infrared triplet is the spectrum synthesis itself. Here we adopt the same
atomic line parameters as specified in Table\,1 of Paper~I, and perform
detailed 3D non-LTE line formation calculations with the
\LINFOR\,\footnote{\url{http://www.aip.de/Members/msteffen/linfor3d}} package, 
which can handle both 3D and
1D model atmospheres. As in the case of the \NLTETD\ calculations of the
departure coefficients, we reduced the horizontal resolution of the 3D model 
atmosphere by choosing only every third grid point in $x$ and $y$-direction. 
Again, we conducted experiments confirming that this reduction does not 
influence the final results (see also Appendix~\ref{app:skipping}).
   
\LINFOR\ computes the LTE populations of the upper and lower levels involved 
in the triplet transitions via Saha-Boltzmann statistics using the \IONOPA\
package\footnote{Since this package is also utilized in \NLTETD, the 
calculations of departure coefficients and spectrum synthesis share
consistent LTE number densities and continuum opacities.}. From this, 
\LINFOR\ evaluates the LTE continuum and line opacities, 
$\kappa_{\rm LTE}^{\rm cont}$ and $\kappa_{\rm LTE}^{\rm line}$. Then it uses
the previously derived departure coefficients to compute the 
NLTE line opacity $\kappa_{\rm NLTE}^{\rm line}$ and line source function 
$S_{\rm NLTE}^{\rm line}$ (assuming complete redistribution), according to

\begin{equation}\label{eq:kappa_NLTE}
\frac{\kappa_{\rm NLTE}^{\rm line}}{\kappa_{\rm LTE}^{\rm line}} =
     b_{\rm l} \,\frac{\exp(\frac{h \nu}{kT}) - (b_{\rm u}/b_{\rm l}) }
           {\exp(\frac{h\nu}{kT})-1}
\end{equation}

\noindent and
   
\begin{equation}\label{eq:source_NLTE}
\frac{S_{\rm NLTE}^{\rm line}}{B_\nu} =
      \frac{\exp(\frac{h\nu}{kT})-1}
           {(b_{\rm l}/b_{\rm u}) \,\exp(\frac{h \nu}{kT})-1}\, ,
\end{equation}
where $B_\nu$ is the Planck function, and $b_{\rm l}$ and $b_{\rm u}$ are the 
departure coefficients of the lower and upper level of the line transition, 
respectively \citep[see, e.g.,][Chapter 4-1]{mihalas78}. 
Note that both LTE and NLTE line opacities are corrected for stimulated 
emission. We recall that we assume 
all transitions of the oxygen 777\,nm triplet to share identical departure 
coefficients $b_{\rm l}$ and $b_{\rm u}$.
The local line profiles are now represented by Voigt profiles, taking  
into account thermal broadening combined with van der Waals and natural
line broadening\footnote{Stark broadening is ignored here as in Paper~I.}; 
differential Doppler shifts along the line of sight due to the 3D 
hydrodynamical velocity field are considered as well.

Using these ingredients, the radiative transfer equation is integrated
on long characteristics to obtain the emergent radiative intensities as
a function of wavelength. The intensity spectra are calculated for two sets 
of inclination angles $\mu$, which represent the WCLC and \citet{PAK09}
observations (see Sect.\,\ref{sect:solar-obs} or Table~\ref{table:EWs_all}), 
in each case for four azimuthal directions (0, $\pi/2$, $\pi$,
3/2\,$\pi$). The final intensity spectrum for each $\mu$ is obtained by
averaging over all surface elements, and to improve the statistics,
over all azimuthal directions. For comparison with the solar flux spectrum, we
computed the integral over $\mu$ of the $\mu$-weighted intensity spectra using
cubic splines  to derive the disk-integrated flux profile.

We finally compute a grid of synthetic line profiles using the corresponding 
grid of departure coefficients described in the previous section. Each 
component of the oxygen triplet is synthesized for \SH-values in the range from 
0 to 8/3 (with steps of 1/3), and \logAO-values in the range between $8.65$ 
and $8.83$ (with steps of 0.03\,dex).

\section{Results}
\label{sect:Solar-abu}

Computed synthetic intensity spectra are compared with observations across
the solar disk  to derive \logAO\ and \SH\ simultaneously. For this
task, we adopt two different methodologies in which we compare theoretical and
observed line profiles and line equivalent widths, respectively. In this 
section, we describe the fitting procedures and their individual results.

\subsection{Line profile fitting}
\label{sect:proffit}

Here we compare observed and theoretical intensity line profiles and
search for the best-fitting \logAO -\SH\ combination. This task utilizes 
the grid of theoretical NLTE line profiles described in 
Sect.\,\ref{sect:specsyn}) and searches for the \logAO -\SH\ combination
that provides the best fit of the intensity profiles of a single triplet 
component simultaneously at all $\mu$-angles.

\begin{figure*}[ht]
 \centering
 \mbox{\includegraphics[width=17cm,clip=true, trim=0 10mm 0 5mm]
 {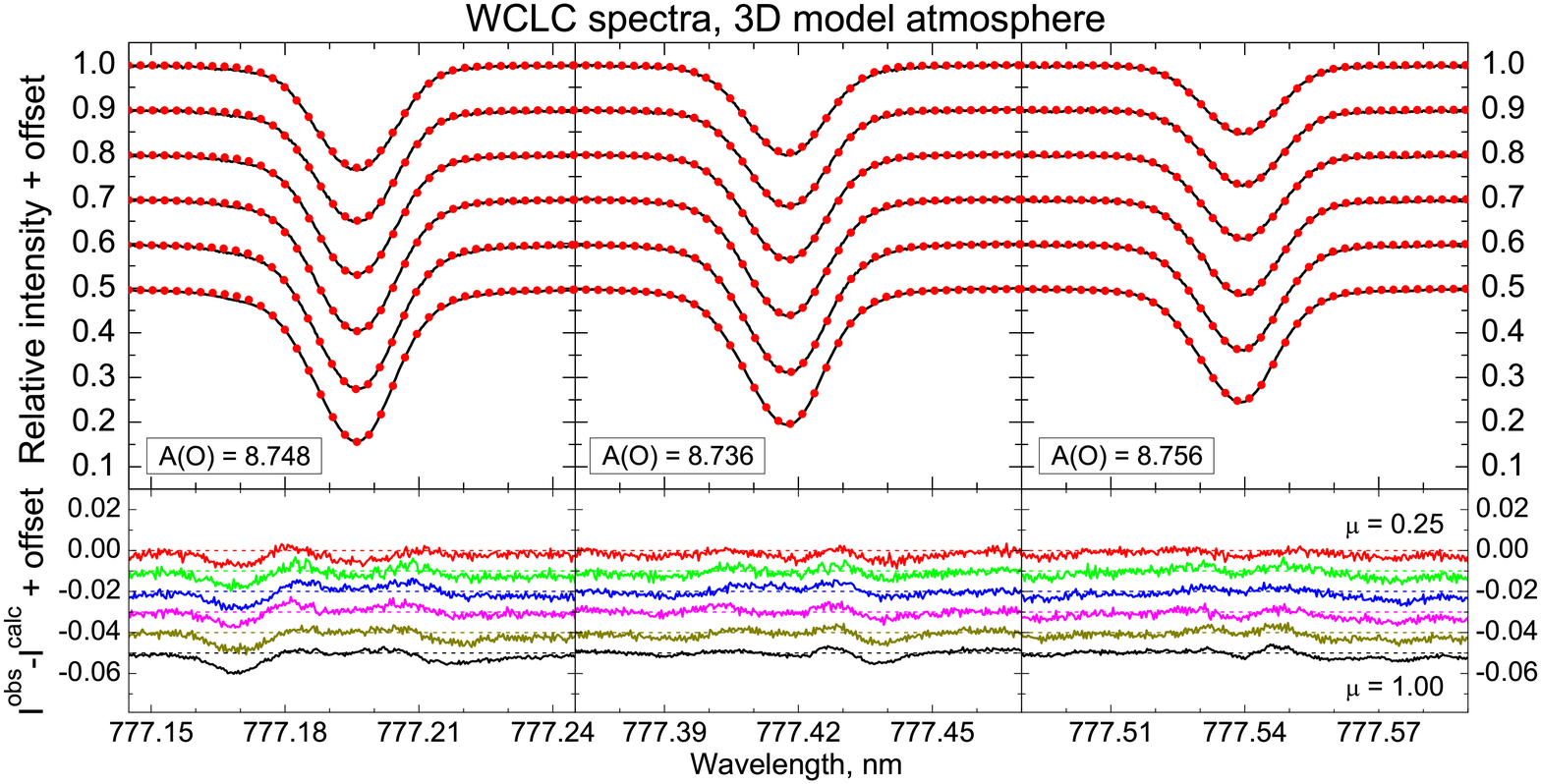}}
 \mbox{\includegraphics[width=17cm,clip=true, trim=0 10mm 0 5mm]
 {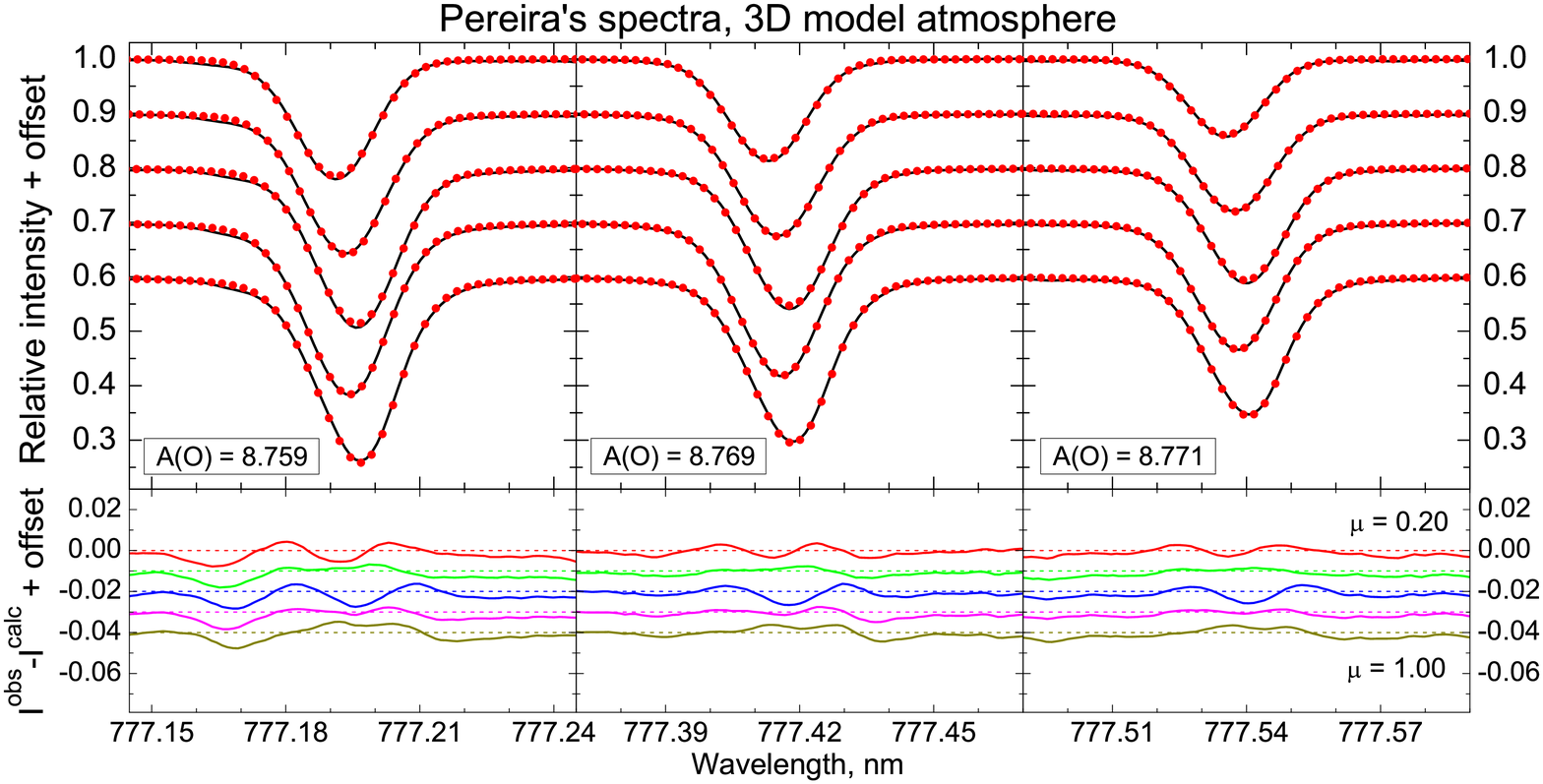}}
\caption{Best-fitting non-LTE line profiles computed with the 3D model
  atmosphere for the WCLC (top) and Pereira (bottom) spectra. The top 
  panels show the observed spectrum (black lines) and their theoretical
  best-fitting counterparts (red dots) for the different $\mu$-angles,
  $\mu$ increasing from top (limb) to bottom (disk center). For clarity, 
  a vertical offset of 0.1 was applied between consecutive $\mu$-angles. 
  The legend shows the best-fitting \logAO\ value for each of the three 
  triplet lines. The bottom panels show the difference of the 
  normalized intensites $I^{\rm obs}-I^{\rm calc}$. An offset of 0.01 was 
  applied between the different $\mu$-angles ($\mu$ increasing downwards).}
\label{fig:ctlProf3D}
\end{figure*}

\begin{table*}[!ht]
\caption {Results of non-LTE line profile fitting with various model 
atmospheres: \logAO, \SH-values, and reduced $\chi^2$ of the best fit
for two different observed spectra (WCLC and Pereira 2009); the difference 
between \logAO\ derived from the two different spectra is given in Col.\,(4).
For the individual triplet components, we give the formal fitting errors
of \logAO\ and \SH\ (only if \SH $>$ 0), while the errors of the mean values
correspond to the standard deviation of the individual lines.}
\label{table:profit}
\centering
\begin{tabular}{|c||c c|c|c c|c c|}
\hline\noalign{\smallskip}
                & \multicolumn{2}{|c|}{\logAO} & ${\rm}\Delta$\logAO &
\multicolumn{2}{|c|}{\SH} & \multicolumn{2}{|c|}{$\chi_{\rm red}^2$}\\
\hline\noalign{\smallskip}
Data set: & WCLC              & Pereira           &        & WCLC            & Pereira         & WCLC  & Pereira  \\
\hline\noalign{\smallskip}
$\lambda$,\,nm & \multicolumn{7}{|c|}{3D model atmosphere}\\ 
\hline
777.2    & $8.748 \pm 0.002$ & $8.759 \pm 0.004$ & -0.011 & $1.44 \pm 0.03$ & $1.64 \pm 0.07$ & 1.654  & 1.794 \\
777.4    & $8.735 \pm 0.002$ & $8.769 \pm 0.004$ & -0.033 & $1.22 \pm 0.03$ & $1.67 \pm 0.07$ & 1.073  & 0.853 \\
777.5    & $8.756 \pm 0.002$ & $8.771 \pm 0.003$ & -0.015 & $1.58 \pm 0.04$ & $1.85 \pm 0.07$ & 1.556  & 1.157 \\
Mean     & $8.747 \pm 0.010$ & $8.766 \pm 0.006$ & -0.019 & $1.41 \pm 0.19$ & $1.72 \pm 0.12$ & --     & --    \\
\hline\noalign{\smallskip}
$\lambda$,\,nm & \multicolumn{7}{|c|}{\xtmean{3D} model atmosphere, $\vmic = 1.0$\,km/s}\\ 
\hline
777.2    & $8.591$ \hspace{10mm} & $8.641 \pm 0.011$     & -0.050 & $0.00$ & $0.22 \pm 0.05$     & 6.163  & 7.010 \\
777.4    & $8.603$ \hspace{10mm} & $8.609$ \hspace{10mm} & -0.006 & $0.00$ & $0.00$\hspace*{9mm} & 3.985  & 3.522 \\
777.5    & $8.614$ \hspace{10mm} & $8.623 \pm 0.010$     & -0.009 & $0.00$ & $0.03 \pm 0.05$     & 2.718  & 2.358 \\
Mean     & $8.603 \pm 0.011$     & $8.624 \pm 0.016$     & -0.021 & --     & $0.08 \pm 0.12$     & --     & --    \\
\hline\noalign{\smallskip}
$\lambda$,\,nm & \multicolumn{7}{|c|}{HM model atmosphere, $\vmic = 0.8$\,km/s}\\ 
\hline
777.2    & $8.671 \pm 0.004$ & $8.673 \pm 0.008$ & -0.002 & $0.44 \pm 0.03$ & $0.46 \pm 0.06$ & 4.409  & 4.387 \\
777.4    & $8.680 \pm 0.004$ & $8.713 \pm 0.007$ & -0.033 & $0.43 \pm 0.03$ & $0.66 \pm 0.06$ & 3.042  & 2.322 \\
777.5    & $8.747 \pm 0.002$ & $8.759 \pm 0.005$ & -0.012 & $1.02 \pm 0.03$ & $1.18 \pm 0.07$ & 2.497  & 2.006 \\
Mean     & $8.699 \pm 0.041$ & $8.715 \pm 0.043$ & -0.016 & $0.63 \pm 0.34$ & $0.76 \pm 0.37$ & --     & --    \\
\hline\noalign{\smallskip}
$\lambda$,\,nm & \multicolumn{7}{|c|}{HM model atmosphere, $\vmic = 1.2$\,km/s}\\ 
\hline
777.2    & $8.635 \pm 0.005$ & $8.638 \pm 0.010$ & -0.003 & $0.38 \pm 0.03$ & $0.40 \pm 0.06$ & 4.935  & 4.998 \\ 
777.4    & $8.644 \pm 0.005$ & $8.671 \pm 0.007$ & -0.027 & $0.34 \pm 0.03$ & $0.49 \pm 0.06$ & 3.349  & 2.652 \\
777.5    & $8.715 \pm 0.002$ & $8.727 \pm 0.005$ & -0.012 & $0.86 \pm 0.03$ & $1.00 \pm 0.07$ & 2.674  & 2.189 \\
Mean     & $8.664 \pm 0.044$ & $8.679 \pm 0.045$ & -0.015 & $0.53 \pm 0.29$ & $0.63 \pm 0.32$ & --     & --    \\
\hline\noalign{\smallskip}
$\lambda$,\,nm & \multicolumn{7}{|c|}{\LHD\ model atmosphere, $\mlp\ = 1.0$, $\vmic = 0.8$\,km/s}\\ 
\hline
777.2    & $8.608$ \hspace{10mm} & $8.607$ \hspace{10mm} &  0.001 & 0.0             & 0.0             & 12.477 & 14.021 \\
777.4    & $8.616$ \hspace{10mm} & $8.620$ \hspace{10mm} & -0.004 & 0.0             & 0.0             &  8.661 &  8.090 \\
777.5    & $8.623$ \hspace{10mm} & $8.625$ \hspace{10mm} & -0.002 & 0.0             & 0.0             &  4.973 &  4.689 \\
Mean     & $8.616  \pm 0.008$    & $8.617  \pm 0.010$    & -0.001 & --              & --              & --     & --     \\
\hline\noalign{\smallskip}
$\lambda$,\,nm & \multicolumn{7}{|c|}{\LHD\ model atmosphere, $\mlp = 1.0$, $\vmic = 1.2$\,km/s}\\ 
\hline
777.2    & $8.581$ \hspace{10mm} & $8.580$ \hspace{10mm} &  0.001 & 0.0             & 0.0             & 14.064 & 15.881 \\
777.4    & $8.593$ \hspace{10mm} & $8.597$ \hspace{10mm} & -0.004 & 0.0             & 0.0             &  9.737 &  9.264 \\
777.5    & $8.604$ \hspace{10mm} & $8.606$ \hspace{10mm} & -0.002 & 0.0             & 0.0             &  5.548 &  5.310 \\
Mean     & $8.593  \pm 0.011$    & $8.594  \pm 0.013$    & -0.001 & --              & --              & --     & --     \\
\hline
\end{tabular} 
\end{table*}

\subsubsection{Fitting procedure}
\label{sect:proffit_0}
We employ the Levenberg-Markwardt least-squares fitting
algorithm \citep{Markwardt2009}, as implemented in IDL procedure MPFIT, to
find the minimum \chisq\ in the four-dimensional parameter space defined by
\logAO, \SH, $v_{\rm b}$, and $\Delta\lambda$. Here $v_{\rm b}$ is the FWHM
(in velocity space) of a Gaussian kernel, which is used to apply some extra
line broadening to the synthetic spectra to account for instrumental
broadening and, in the case of 1D model atmospheres, macroturbulence; 
$\Delta\lambda$ is a global wavelength shift applied to the synthetic
spectrum  to compensate for 
potential imperfections of the wavelength calibration, uncertainties 
in the laboratory wavelengths, and systematic component- and $\mu$-dependent 
line shifts. Test runs with an additional free fitting parameter for the continuum placement resulted in unsatisfactory fits. We therefore rely on the 
correct normalization of the observed spectra and consider the continuum
level as fixed. 

In a first sweep, we find the minimum \chisq\ for each triplet component
$k$ and each $\mu$-angle individually. Here \chisq\ is defined as
\begin{equation}
\label{eq:chi2profit1}
   \chi^2 = \sum_{i=1}^{n} 
   \frac{\left(I_i^{\rm obs}-I_i^{\rm calc}\right)^2}{\sigma_i^2}\, ,
\end{equation}
where  the sum is taken over all $n$ considered wavelength points of the 
line profile, $I_i^{\rm obs}$ and $I_i^{\rm calc}$ is the intensity of
observed and synthetic normalized spectrum, respectively, 
and $\sigma_i$ is the pixel-dependent observational uncertainty of 
the spectrum to be fitted. For simplicity, we have adopted  a fixed continuum signal-to-noise ratio of 
$S/N = 500$ for both the 
WCLC and the Pereira spectra, and computed $\sigma_i$ simply as $\sigma = (S/N)^{-1}$, 
independent of wavelength point $i$ and inclination angle $\mu$.

An obvious blend in the blue wing of the strongest triplet component 
at $\lambda\, 777.2$\,nm is excluded from the relevant wavelength range
by masking a window between $-374$ and $-214$\,m\AA\ from the line center,
setting $1/\sigma_i=0$ for these pixels.

The result of this \chisq\ minimization procedure is a set of best-fit 
parameters \logAO, \SH, $v_{\rm b}$, $\Delta\lambda$ for each ($k, \mu$).
These intermediate results are only used for two purposes: (i) for fixing 
the \emph{\textup{relative}} wavelength shift of the line profiles as a function of 
$\mu$ for each triplet component $k$; and (ii) to `measure' the equivalent 
width of each individual observed line profile via numerical integration 
of the best-fitting synthetic line profile over a fixed wavelength 
interval of $\pm2$\,\AA. The measured equivalent widths are used later 
for the equivalent width fitting method as described in 
Sect.\,\ref{sect:EWfit_1}.

We note that the oxygen abundance obtained from fitting the individual 
line profiles is relatively high, $8.78 < \logAO < 8.87$, where the
required oxygen abundance tends to increase towards the limb. At the same time,
the corresponding \SH-values are large, too ($\SH \ga 3.0$). However, we
do not consider these results to be relevant because the line profile 
at a given single $\mu$ can be reproduced almost equally well with different 
combinations of the fitting parameters \logAO, \SH, and $v_{\rm b}$, which are
tightly correlated. Small imperfections in the theoretical and / or observed
line profiles can thus easily lead to ill-defined solutions.
This degeneracy is only lifted when fitting the profiles at all $\mu$-angles
simultaneously.

In the second sweep, we fit the line profiles at all $\mu$-angles ($\mu_j,
j=1\ldots m$) simultaneously, but separately for each of the three triplet 
components. In this case, \chisq\ is defined as
\begin{equation}
\label{eq:chi2profit2}
   \chi^{2} = \sum_{j=1}^{m}\sum_{i=1}^{n} 
   \frac{\left(I_{i,j}^{\rm obs}-I_{i,j}^{\rm calc}\right)^2}{\sigma_{i,j}^2}\, ,
\end{equation}
where $m$ is the number of considered $\mu$-angles ($\mu \ga 0.2$). 
Again, $\sigma_{i,j}$ is taken to be constant, 
$\sigma_{i,j}=\sigma = (S/N)^{-1}$, as explained above, and the blend in 
the blue wing of the strongest triplet line at $\lambda\,777.2$\,nn is 
masked as before. Treating the triplet components separately has 
the advantage of allowing a consistency check and providing an independent 
error estimate of the derived oxygen abundance.

As in the first sweep, we consider \logAO, \SH, and $v_{\rm b}$ as free 
fitting parameters, but instead of a single global $\Delta\lambda$,
we introduce two global line shift parameters, $\Delta\lambda_1$ and 
$\Delta\lambda_2$, to allow for the most general case of
a $\mu$-dependent line shift correction, computed as 
\mbox{$\Delta\lambda(\mu) = \Delta\lambda_0(\mu) + \Delta\lambda_1 + 
\Delta\lambda_2\,(1-\mu)$}, where $\Delta\lambda_0(\mu)$ is the individual
wavelength shift determined in the first sweep for the respective $\mu$-angle.
It turns out, however, that the results of the global fitting are 
insensitive to the detailed treatment of the line shifts.

\subsubsection{Fitting results for the WCLC and Pereira spectra}
\label{sect:proffit_1}
Fig.~\ref{fig:ctlProf3D} shows the results of the simultaneous line profile
fitting with the 3D-NLTE synthetic line profiles of the WCLC and Pereira
spectrum, respectively. In general, the agreement
between observation and model prediction is very good; the differences
never exceed 1\% of the continuum intensity and usually remain at a much lower
level. Both the line wings and  line core are represented very well at all 
inclination angles. The very satisfactory match we obtain between 
theory and observation suggests that both the velocity field and the thermal
structure of our model atmosphere are realistic in the range of optical depths 
sampled by the \ion{O}{i} IR triplet lines for inclination angles $\mu \ga 0.2$.
In addition, the accurate reproduction of the observed line profiles suggests
that our NLTE line formation technique is adequate, despite the possible 
shortcomings regarding the modeling of hydrogen collisions.
   
The corresponding fits obtained with the 1D hydrostatic models are collected 
in Appendix~\ref{app:1Dprofiles}. In general, the line profiles computed with 
the 1D hydrostatic model atmospheres result in a less satisfactory agreement 
with the observations. For example, given a single \logAO\ and \SH\ value, 
the \xtmean{3D} and 1D \LHD\ models fail to simultaneously account for the 
line core and wings at $\mu = 1$, and predict a much too weak line 
core close to the limb (see Figs.\,\ref{fig:ctlProf3Dmean_mic10} and 
\ref{fig:ctlProf1D_mic12}). The situation with the HM model 
(Fig.\,\ref{fig:ctlProfHM_mic12}) is better, but the line profile fits are 
still clearly inferior to those computed with the 3D model atmosphere. It is 
worth mentioning that, for all 1D model atmospheres, changing the 
($\mu$-independent)
microturbulence parameter only changes the value of the best-fitting \logAO\ 
and \SH, but does not improve the quality of the fits. 

In Table~\ref{table:profit}, we summarize the main results derived from 
fitting the line profiles simultaneously for all $\mu$-angles, separately
for each of the three triplet components, and for both the WCLC and the Pereira 
observed spectra. Columns (2) and (3) list the best-fitting oxygen abundance
\logAO\ and their formal fitting errors for WCLC and Pereira, respectively, 
and Col.\,(4) shows the difference between \logAO\ derived from the two 
different sets of spectra. Columns (5) and (6) give the corresponding 
results for \SH, while the last two columns show the reduced \chisq\ of 
the best fit for the two sets of spectra, respectively, defined as 
$\chi_{\rm red}^2=\chisq/(n\times m - p$), where $p = 5$ is the number 
of free fitting parameters. The fact that $\chi_{\rm red}^2$ is of the order of 
unity means that the fit provided by the synthetic spectra is good (at least
for the 3D model), given the assumed uncertainty of the observed spectral
intensity of $\sigma_i = 1/500$. However, only relative \chisq-values are 
considered meaningful for judging the quality of a fit.

The formal fitting errors of \logAO\ and \SH\ given in
Table~\ref{table:profit} are computed as the square root of the corresponding 
diagonal elements of the covariance matrix of the fitting parameters, which 
is provided by the MPFIT procedure. The fact that these fitting errors are
relatively small shows that the solution found by the fitting procedure is
well defined. However, these formal errors do not account for possible 
systematic errors (e.g., due to imperfect continuum placement, unknown weak
blends, insufficient time averaging, etc.), and hence the true errors are
likely to be larger. In the case of the \xtmean{3D} and \LHD\ models, the formal
errors are meaningless because the best fit would require \SH$<$0 (see
below), and the constraint \SH$\ge 0$ essentially enforces \SH=0.

As shown in Table~\ref{table:profit}, the oxygen abundances derived from 
the two sets of observed spectra generally agree within 0.03\,dex or better, 
which is of the same order of magnitude as the line-to-line variation of 
\logAO. The WCLC spectra produce systematically lower abundances. We 
attribute this to the systematically stronger lines at low $\mu$ compared to 
the \citet{PAK09} data (cf.\ Fig.\,\ref{fig:W_obs_ctl} below), which implies
a reduced gradient $W(\mu)$, and, in turn, a lower \SH\ and \logAO, as 
explained in more detail in Sect.\,\ref{sect:discussion2}. In the case of 
the 3D model atmosphere, we consider the agreement in the derived \logAO\ 
between the different triplet lines and between the two sets of spectra as 
satisfactory.

In case of 1D \LHD\ model atmospheres, the agreement between the best-fit 
parameters of the two spectra is nearly perfect, but this is again related 
to the fact that \SH\ is forced to 0. The resulting fits are 
not good (high $\chi_{\rm red}^2$), particularly at the lowest $\mu$-angle, as
mentioned above. The derived \logAO\ are significantly lower (by about 
$-0.15$\,dex) than for the 3D case. Similar conclusions can be drawn for 
the \xtmean{3D} model.
The non-LTE oxygen abundance obtained with the 1D semi-empirical HM model 
atmosphere lies somewhere in between the 3D and the 1D \LHD\ results.
While \SH\ is positive, the line profiles are poorly fitted, both at disk 
center and near the limb (see Fig.\,\ref{fig:ctlProfHM_mic12}), and the 
line-to-line variation of \logAO\ is rather high.

As will be discussed below in Sect.\,\ref{sect:discussion2}, all 1D results 
are systematically biased towards low \SH\ and low \logAO\ values. 
In the following, we shall therefore ignore the 1D results for the 
determination of the solar oxygen abundance.

To estimate the solar oxygen abundance from the profile fitting
method, we compute, separately for each of the two data sets, the average
abundance, $\left<{\logAO}\right>$, as the unweighted mean over the
best-fitting \logAO\ values of the three triplet lines. These mean oxygen
abundances are listed in Table~\ref{table:profit} in the rows marked `Mean',
together with their errors, which in this case are defined by the
standard deviation of the individual data points from the three triplet 
components.

From the 3D-NLTE synthetic spectra we find: $\left<{\logAO}\right> = 
8.747 \pm 0.010$\,dex (WCLC spectra) and $\left<{\logAO}\right> = 
8.766 \pm 0.006$\,dex (Pereira's spectra). We note that the $1\,\sigma$ 
error bars of the two results fail (by a narrow margin) to overlap, 
indicative of unaccounted systematic errors. We finally 
derive a two-spectra average of $\left<\left<{\logAO}\right>\right> 
= 8.757 \pm 0.01$\,dex ($\left<\left<{\SH}\right>\right>=1.6 \pm 0.2$).

\subsubsection{Fitting of disk-center intensity and flux line profiles}
\label{sect:proffit_2}
As an independent consistency check, we  also applied our line profile fitting
procedure  to the \citet{neckelobs} FTS spectra. In this case, we
simultaneously fit the solar disk-center (Neckel Intensity) and
disk-integrated (Neckel Flux) \ion{O}{i} IR triplet line profiles.
The blue wing of the $777.2$\,nm line is masked as before.
The synthetic flux spectrum is computed by integration over $\mu$ of the 
$\mu$-weighted intensity spectra available at the five inclinations 
defined by the Pereira spectra. Subsequently, the 
flux spectrum is broadened with the solar synodic projected rotational 
velocity $v\,\sin i = 1.8$\,km/s. The fitting parameters are again \logAO, \SH, 
instrumental broadening $v_{\rm b}$, and the individual line shifts
$\Delta\lambda_1$ and $\Delta\lambda_2$ for intensity and flux profile, 
respectively.

Using the 3D non-LTE synthetic line profiles, we obtain an average 
oxygen abundance of \logAO\ $=8.80 \pm 0.008$, where this error 
refers to the formal fitting error; the line-to-line scatter is 
much smaller, $\la 0.001$\,dex. The best-fitting mean \SH\ value is 
$2.7 \pm 0.25$. As before, the
quality of the fits is very satisfactory, with $\chi_{\rm red}^2 = 0.81$, 
$0.84$, and $1.13$ for the three triplet components at $\lambda$\,$777.2$,
$777.4$, and $777.5$\,nm, respectively.

Both \logAO\ and \SH\
are significantly higher than inferred from the center-to-limb variation
of the line profiles in the WCLC and Pereira spectra. We have to conclude 
that neither the formal fitting errors nor the line-to-line variation of
\logAO\ represent a valid estimate of the true uncertainty of the derived
oxygen abundance. Rather, systematic errors must be the dominating factor.
We shall demonstrate in the next Section that the result of  
simultaneously fitting the intensity and flux profiles is very sensitive to 
small errors in measuring or modeling the relative line strengths in the
disk-center and integrated disk spectra. We therefore consider the oxygen
abundance derived from fitting the spectra at several $\mu$-angles
simultaneously as more reliable.

\subsection{Equivalent width fitting} 
\label{sect:EWfit}

Another possible methodology for estimating the solar oxygen abundance is
based on the comparison of the observed equivalent widths (EWs) and their 
center-to-limb variation with their theoretical counterparts. Matching the 
EW of the lines is admittedly less precise than line profile fitting: it 
is possible that synthetic and observed EWs match perfectly, while the line 
profiles exhibit significantly different shapes. However, EW-fitting has 
the advantage of being simpler: there are only two free fitting parameters, 
\logAO\ and \SH. The additional fitting parameters that were necessary in 
the case of line profile fitting, related to extra line broadening and line 
shifts, have no influence on the equivalent width and thus are irrelevant 
for the EW fitting procedure.
   
\begin{table}[!ht]
\caption {EWs of the \ion{O}{i} IR triplet as a function of $\mu$ 
measured from the different spectra with our line profile fitting 
method (this work); for the Pereira spectra, we also give the original 
measurements taken from \citet{PAK09}.}
\label{table:EWs_all} 
\centering
\begin{tabular}{|c|c|c|c|}
\hline 
      & \multicolumn{3}{c|}{Equivalent width (pm)} \\
\hline\noalign{\smallskip} 
      & \ion{O}{i}\,7772 & \ion{O}{i}\,7774 & \ion{O}{i}\,7775 \\ 
\hline\noalign{\smallskip} 
$\mu$ & \multicolumn{3}{c|}{Spectrum: \citet{neckelobs}, EW: this work}\\ 
\hline\noalign{\smallskip} 
1.00  & $8.73 \pm 0.050$  & $7.60 \pm 0.044$  & $6.01 \pm 0.035$ \\
flux  & $7.54 \pm 0.043$  & $6.52 \pm 0.038$  & $5.07 \pm 0.029$ \\
\hline\noalign{\smallskip} 
$\mu$ & \multicolumn{3}{c|}{Spectrum: WCLC, EW: this work}\\ 
\hline\noalign{\smallskip} 
1.00  & $8.73 \pm 0.040$  & $7.51 \pm 0.035$  & $5.95 \pm 0.027$ \\
0.87  & $8.27 \pm 0.038$  & $7.13 \pm 0.033$  & $5.63 \pm 0.026$ \\
0.66  & $7.33 \pm 0.034$  & $6.34 \pm 0.029$  & $5.02 \pm 0.023$ \\
0.48  & $6.54 \pm 0.030$  & $5.62 \pm 0.026$  & $4.36 \pm 0.020$ \\
0.35  & $5.94 \pm 0.027$  & $5.08 \pm 0.023$  & $3.79 \pm 0.017$ \\
0.25  & $5.44 \pm 0.025$  & $4.63 \pm 0.021$  & $3.40 \pm 0.016$ \\
\hline\noalign{\smallskip} 
$\mu$ & \multicolumn{3}{c|}{Spectrum: \citet{PAK09}, EW: this work}\\ 
\hline\noalign{\smallskip} 
1.00  & $8.63 \pm 0.060$ & $7.55 \pm 0.052$ & $5.99 \pm 0.041$ \\
0.816 & $8.08 \pm 0.056$ & $7.03 \pm 0.049$ & $5.54 \pm 0.038$ \\
0.608 & $7.22 \pm 0.050$ & $6.25 \pm 0.043$ & $4.85 \pm 0.034$ \\
0.424 & $6.33 \pm 0.044$ & $5.46 \pm 0.038$ & $4.16 \pm 0.029$ \\
0.197 & $4.94 \pm 0.034$ & $4.16 \pm 0.029$ & $3.02 \pm 0.021$ \\
\hline\noalign{\smallskip} 
$\mu$ & \multicolumn{3}{c|}{Spectrum and EW: \citet{PAK09}}\\ 
\hline\noalign{\smallskip} 
1.00  & $8.48 \pm 0.09 $ & $7.44 \pm 0.09 $ & $5.90 \pm 0.09 $ \\
0.816 & $8.00 \pm 0.14 $ & $6.96 \pm 0.15 $ & $5.47 \pm 0.11 $ \\
0.608 & $7.21 \pm 0.09 $ & $6.21 \pm 0.09 $ & $4.83 \pm 0.09 $ \\
0.424 & $6.38 \pm 0.10 $ & $5.44 \pm 0.10 $ & $4.17 \pm 0.10 $ \\
0.197 & $5.09 \pm 0.10 $ & $4.22 \pm 0.33 $ & $3.08 \pm 0.28 $ \\
\hline
\end{tabular} 
\end{table}

\begin{figure}[!htb]
 \centering
 \mbox{\includegraphics[bb=20 20 340 260,width=9cm]{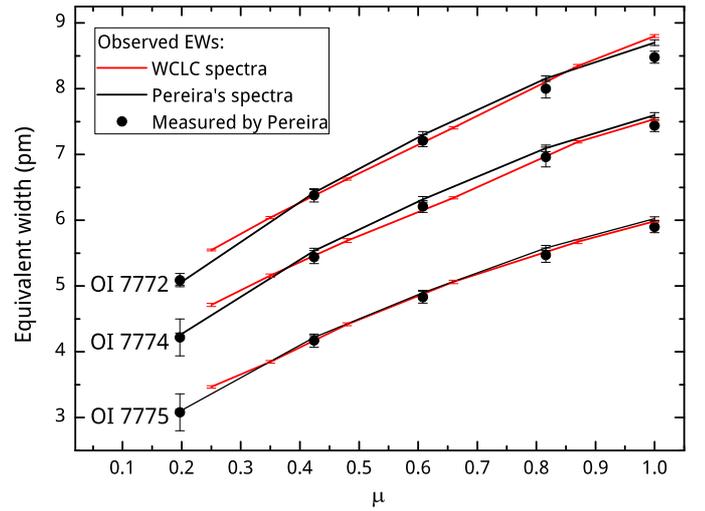}}
 \caption{Equivalent widths as measured in this work from WCLC and Pereira's 
  spectra (red and black lines, respectively), compared to the measurements 
  taken from the work of \citet{PAK09} (black dots). See 
  Table~\ref{table:EWs_all} for more details.}
 \label{fig:W_obs_ctl}
\end{figure}

\begin{figure*}[!htb]
 \centering
 \mbox{\includegraphics[width=17cm]{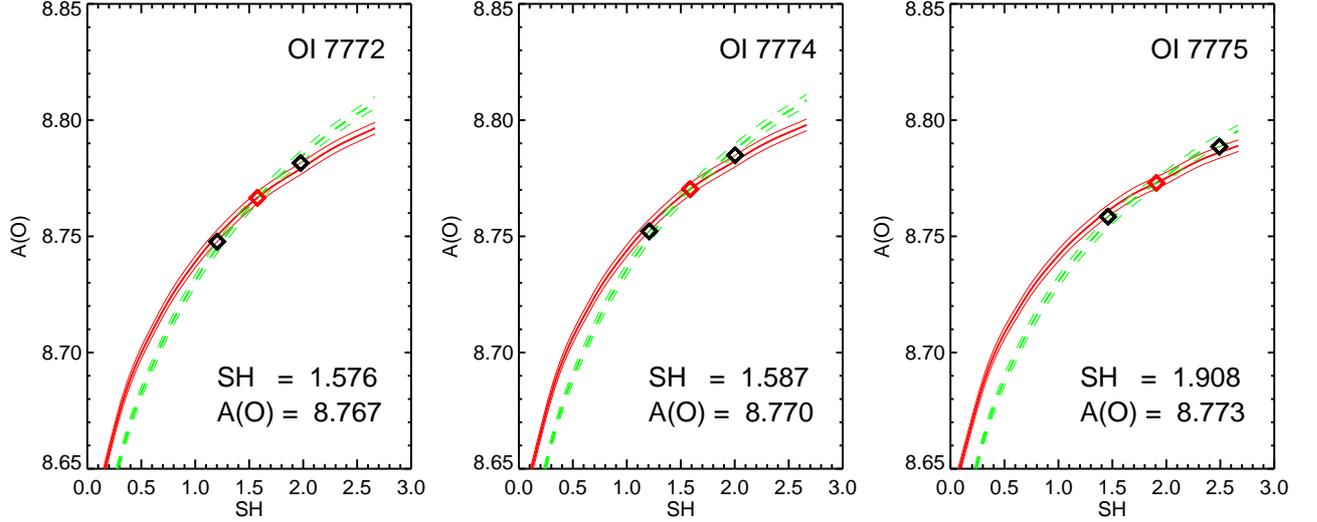}}
 \caption{Lines of constant equivalent width in the \SH\ -- \logAO\ plane, 
  as computed from the grid of 3D non-LTE synthetic line profiles for the
  three triplet components.
  The thick (red) solid line refers to the EW measured in the Neckel
  disk-center spectrum, while the thick (green) dashed line is for the
  measured EW in the Neckel disk-integrated (flux) spectrum. The solution 
  (\SH, \logAO) given in the legend is defined by the intersection point of
  the two lines (red diamond). The black diamonds define the uncertainty
  range of the solution, assuming an EW measurement error of $\approx
  0.5$\%, as indicated by the thin lines offset from the original 
  \SH\ -- \logAO\ relations by $\pm\Delta\logAO=0.0025$.}
 \label{fig:W_fit_neckel}
\end{figure*}

\begin{figure*}
\centering
\mbox{\includegraphics[bb=0 0 1201 400,width=17cm,clip=true]
{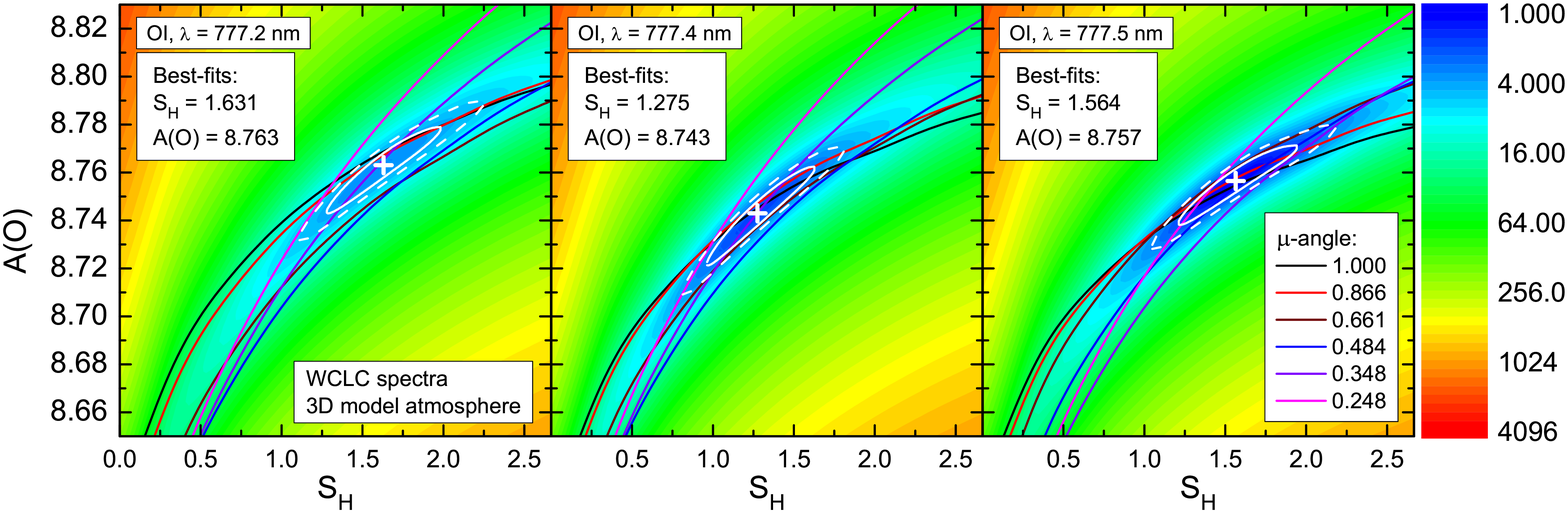}}
\mbox{\includegraphics[bb=0 0 1201 400,width=17cm,clip=true]
{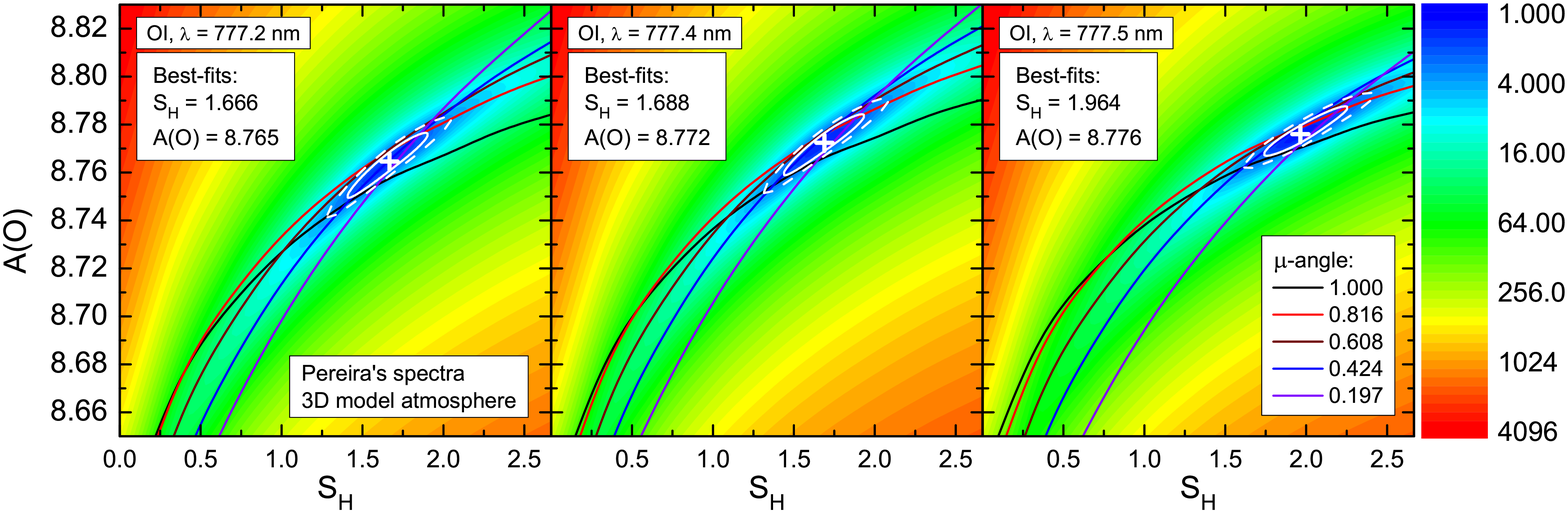}}
 \caption{Maps of $\chi^2$ (color-coded) in the \SH\ -- \logAO\ plane, 
  indicating the quality of the simultaneous fit over all $\mu$-angles 
  of observed and computed 3D non-LTE EW, considering each of the three 
  triplet components separately. The location of the minium $\chi^2$ is
  marked by a white plus sign, while the white ellipses bound the regions 
  where $\chi^2 - \chi_{\rm min}^2 < 1$ (solid) and $2.3$ (dashed). The latter
  corresponds to the simultaneous confidence region of \SH\ and \logAO\
  containing $68.3$\,\% in the case of normally distributed data.
  Contour lines indicating where 
  $W_{\rm obs} = W_{\rm calc}$ are superimposed for each $\mu$-angle. 
  The top and bottom panels show the results for the WCLC and Pereira 
  spectra, respectively. Line identification and best-fit parameters 
  are given at the top left of each panel.}
 \label{fig:EWcontours}
\end{figure*}

\subsubsection{Equivalent width measurements}
\label{sect:EWfit_1}
For a given model atmosphere, a grid of theoretical equivalent widths,
$W_{\rm calc}(\mu)$, is computed as a function \logAO\ and \SH\ by integrating 
the non-LTE synthetic line profile over a wavelength interval of 
$\pm 2$\,\AA, separately for each triplet component. Here the inclination 
angle $\mu$ covers the values for both observed spectra (WCLC and Pereira).
In addition, the equivalent widths of the disk-integrated line profiles were
obtained by appropriate $\mu$-integration, denoted as
$\overline{W}_{\rm calc}$ in the following.

The EWs of the observed line profiles, $W_{\rm obs}(\mu)$,  
$\overline{W}_{\rm obs}$, are obtained as a by-product of the 3D non-LTE
line profile fitting (see Sect.\,\ref{sect:proffit}): $W_{\rm obs} = 
W_{\rm calc}(\logAO^\ast, \SH^\ast)$, where $\logAO^\ast$ and $\SH^\ast$
are the best-fit parameters for the respective line profile,
obtained by fitting each $\mu$-angle individually.
This method of EW measurement ensures an optimal consistency between 
observed and theoretical equivalent widths.
The errors of the observed EWs, $\sigma_{\rm  o}$, were estimated from the 
formal errors of \logAO\ and \SH\, for the best-fitting theoretical line 
profile. The EWs of the \ion{O}{i} IR triplet measured in this way 
are listed in Tab.\,\ref{table:EWs_all} for all spectra;
for the Pereira spectra, we also give the original measurements 
by \citet{PAK09} (bottom). The EWs from both the WCLC and the Pereira
spectra, as collected in Tab.\,\ref{table:EWs_all}, are plotted in 
Fig.\,\ref{fig:W_obs_ctl}.

The EWs derived with our method from the Pereira spectra are
generally slightly larger than those measured by the authors of the original
work. This is because we integrate the
synthetic single line profile over a very broad wavelength window. Since 
the \ion{O}{i} lines have extended
Lorentzian wings, our EWs are systematically higher. For the weakest lines
at $\mu = 0.197$, the impact of the Lorentzian wings is diminished,
and the two measurements are in very close agreement 
(see Table~\ref{table:EWs_all}). We argue that, for the purpose of EW fitting,
our EW measurement method is superior to direct integration of the observed 
line profiles because it ensures consistency of theoretical and observed 
EWs.

\subsubsection{EW fitting of disk-center and full-disk spectra}
\label{sect:EWfit_2}
We first apply the method to determine \SH\ and \logAO\ from the simultaneous
fit of the disk-center and disk-integrated EWs in the Neckel intensity and
flux spectra, respectively. For this purpose, we compute for each of the three
triplet components two grids of equivalent widths, $W$(\SH, \logAO; $\mu$=$1$)
and $W$(\SH, \logAO; flux), from the corresponding sets of 3D non-LTE line
profiles. These grids are then used to construct contour lines in the \SH\ --
\logAO\ plane where $W_{\rm calc} = W_{\rm obs}$, as shown in
Fig.\,\ref{fig:W_fit_neckel}. The thick (red) solid line and the thick (green)
dashed line refers to the EW measured in the Neckel disk-center and
disk-integrated (flux) spectrum, respectively (see
Table~\ref{table:EWs_all}). The slope of the contour line of the flux EW is
somewhat steeper than that for disk-center, indicating that the lines in the
flux spectrum are slightly more sensitive to changes in \SH. The unique
solution (\SH, \logAO) defined by the intersection point of the two lines (red
diamond) is given in the legend of each panel.

Since the angle at which the two lines intersect is rather small, the position
of the intersection point is very sensitive to the exact values of the two
equivalent widths. To quantify the uncertainties, we have also plotted in 
Fig.\,\ref{fig:W_fit_neckel} (thin lines) the contour lines shifted up and 
down by  $\Delta$\logAO = $\pm 0.0025$\, dex, corresponding to errors in EW 
of less than $0.5$\%. The black diamonds show the solutions obtained by 
changing the two EWs by this small amount in opposite directions. With this 
definition of the error bars, we find \logAO\ = $8.77 \pm 0.02$, 
\SH = $1.85 \pm 0.65$.

These results are somewhat lower than the corresponding results obtained from
line profile fitting (Sect.\,\ref{sect:proffit_2}), with a marginal overlap of
the error bars. We conclude that the idea of simultaneously fitting intensity 
and flux spectra works in principle, but is highly sensitive to small 
errors in the measured EWs and to systematic errors in modeling the 
$\mu$ dependence of the oxygen triplet lines. The results obtained with this
method are therefore only considered  a rough indication of the range of 
\logAO\ compatible with our modeling. As we  see below, a more accurate 
determination of the oxygen abundance is possible when fitting simultaneously 
the line profiles measured at different $\mu$-angles.

\begin{table*}
\caption{Results of non-LTE EW fitting with various model 
atmospheres: \logAO, \SH-values, and reduced $\chi^2$ of the best fit
for two different observed spectra (WCLC and Pereira 2009); the difference 
between \logAO\ derived from the two different spectra is given in Col.\,(4).
Results from the \xtmean{3D} and \LHD\ model atmospheres are excluded 
because no minimum in \chisq\ is found in the range \mbox{\SH\ $\ge$ $0$.}}
\label{table:EWfit}
\centering
\begin{tabular}{|c||c c|c|c c|c c|}
\hline\noalign{\smallskip}
                & \multicolumn{2}{|c|}{\logAO} & ${\rm}\Delta$\logAO &
\multicolumn{2}{|c|}{\SH} & \multicolumn{2}{|c|}{$\chi_{\rm red}^2$}\\
\hline\noalign{\smallskip}
Data set: & WCLC              & Pereira           &        & WCLC            & Pereira         & WCLC  & Pereira  \\
\hline
$\lambda$,\,nm & \multicolumn{7}{|c|}{3D model atmosphere}\\ 
\hline
777.2    & $8.763 \pm 0.018$ & $8.765 \pm 0.014$ & -0.002 & $1.63 \pm 0.35$ & $1.67 \pm 0.25$ & 0.831 & 0.500 \\
777.4    & $8.743 \pm 0.021$ & $8.772 \pm 0.013$ & -0.029 & $1.28 \pm 0.33$ & $1.69 \pm 0.25$ & 0.448 & 0.457 \\
777.5    & $8.757 \pm 0.017$ & $8.776 \pm 0.010$ & -0.019 & $1.56 \pm 0.36$ & $1.96 \pm 0.26$ & 0.235 & 0.401 \\
Mean     & $8.754 \pm 0.010$ & $8.771 \pm 0.006$ & -0.017 & $1.49 \pm 0.19$ & $1.77 \pm 0.17$ & -     & -     \\
\hline
$\lambda$,\,nm & \multicolumn{7}{|c|}{HM model atmosphere, $\vmic = 0.8$\,km/s}\\ 
\hline
777.2    & $8.699 \pm 0.030$ & $8.677 \pm 0.031$ & ~0.022 & $0.55 \pm 0.25$ & $0.38 \pm 0.16$ & 2.135 & 0.989 \\
777.4    & $8.689 \pm 0.031$ & $8.705 \pm 0.022$ & -0.016 & $0.42 \pm 0.22$ & $0.49 \pm 0.17$ & 1.414 & 0.477 \\
777.5    & $8.731 \pm 0.022$ & $8.748 \pm 0.014$ & -0.017 & $0.73 \pm 0.26$ & $0.90 \pm 0.19$ & 0.371 & 0.159 \\
Mean     & $8.706 \pm 0.022$ & $8.710 \pm 0.035$ & -0.004 & $0.57 \pm 0.16$ & $0.59 \pm 0.28$ & -     & -     \\
\hline
$\lambda$,\,nm & \multicolumn{7}{|c|}{HM model atmosphere, $\vmic = 1.2$\,km/s}\\ 
\hline
777.2    & $8.670 \pm 0.044$ & $8.650 \pm 0.026$ & ~0.020 & $0.50 \pm 0.21$ & $0.36 \pm 0.08$ & 2.330 & 1.371 \\
777.4    & $8.660 \pm 0.030$ & $8.674 \pm 0.026$ & -0.014 & $0.37 \pm 0.16$ & $0.42 \pm 0.16$ & 1.578 & 0.695 \\
777.5    & $8.705 \pm 0.023$ & $8.721 \pm 0.015$ & -0.016 & $0.65 \pm 0.25$ & $0.80 \pm 0.18$ & 0.420 & 0.208 \\
Mean     & $8.678 \pm 0.024$ & $8.682 \pm 0.036$ & -0.004 & $0.51 \pm 0.14$ & $0.52 \pm 0.24$ & -     & -     \\
\hline
\end{tabular} 
\end{table*}

\begin{figure*}[htb]
 \centering
 \mbox{\includegraphics[width=8.5cm,clip=true]{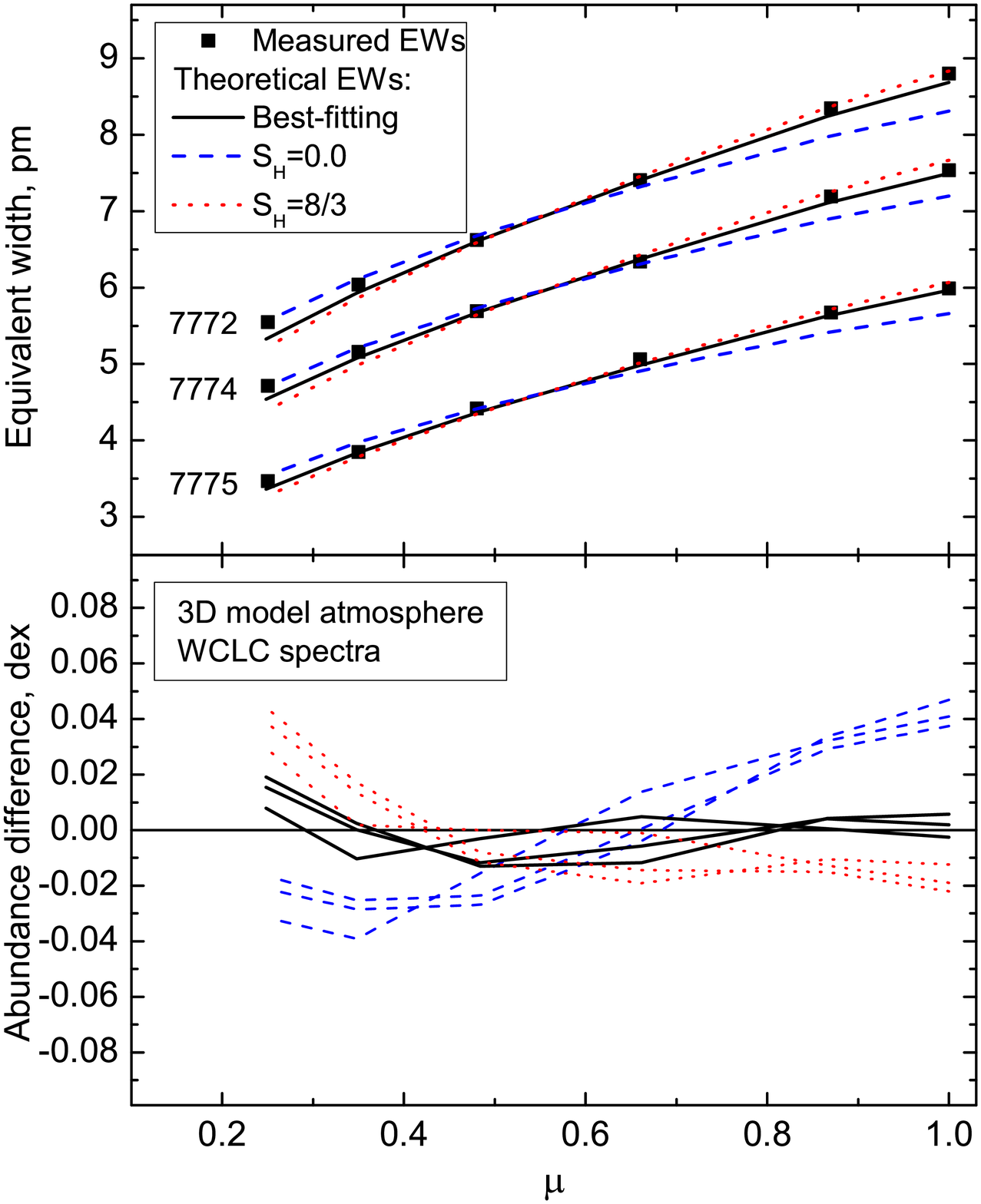}}
 \mbox{\includegraphics[width=8.5cm,clip=true]{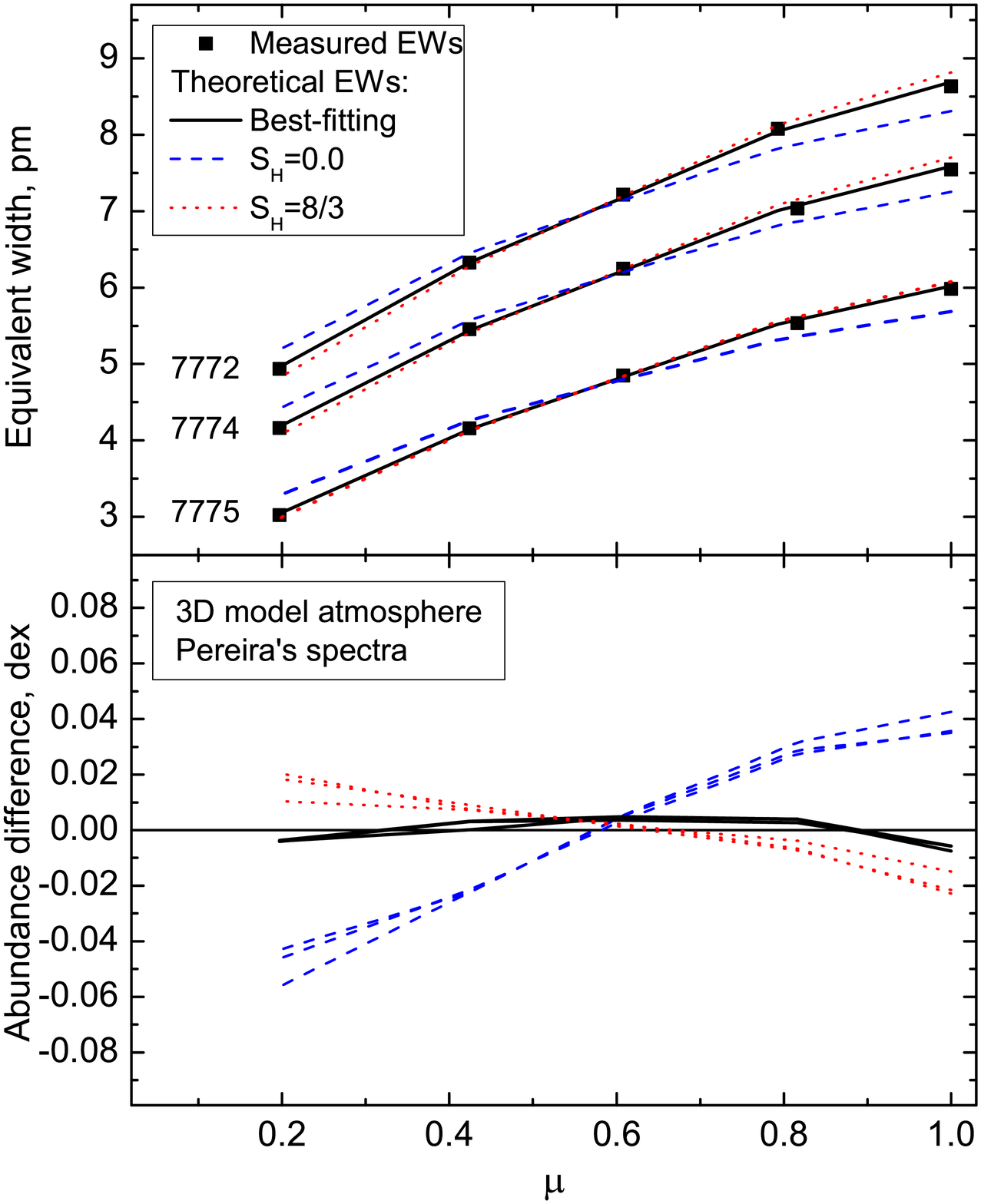}}
 \caption{{\it Top panels}: center-to-limb variation of observed (symbols) 
 and theoretical EWs derived from the 3D model atmosphere for \SH=0 (blue 
 dashed), 8/3 (red dotted) and best-fit value (black solid), shown 
 individually for the three triplet components, with wavelength increasing 
 from top to bottom. For each \SH, the corresponding \logAO\ is defined as 
 the value that minimizes \chisq\ (Eq.\,\ref{eq:chi2ewfit}) at fixed \SH. 
 {\it Bottom panels}:
 abundance difference A$(\mu)$ - $\overline{{\rm A}}$, where A$(\mu)$
 and $\overline{{\rm A}}$ denote \logAO\ obtained from line profile
 fitting at individual $\mu$-angles and at all $\mu$-angles simultaneously. 
 Left and right panels refer to the WCLC and Pereira spectra, respectively.}
 \label{fig:ctlEWs}
\end{figure*}

\subsubsection{EW fitting of the $\mu$-dependent WCLC and Pereira spectra}
\label{sect:EWfit_3}
   
We now determine \SH\ and \logAO\ from the simultaneous fit of  
the EWs at all $\mu$-angles, using the two independent sets of observed
$\mu$-dependent spectra of WCLC and Pereira, respectively. As before, we 
treat the three triplet components separately. In contrast to the method
employed by \cite{PAK09}, we do not assign a special role to the fit
at disk center. The idea of our method is to find the minimum \chisq\ in
the \SH\ -- \logAO\ plane, where \chisq\ is defined as

\begin{equation}
\label{eq:chi2ewfit}
   \chi^{2}(\SH,\logAO) = \sum_{\rm j=1}^{m} 
   \frac{\left(W_j^{\rm obs}-W_j^{\rm calc}(\SH,\logAO)\right)^2}
   {\sigma_{1,j}^2+\sigma_{2,j}^2}\, .
\end{equation}
We sum over the different $\mu$-angles ($j=1\ldots m$); 
$W^{\rm obs}$ and $W^{\rm calc}$ are the measured and computed
equivalent width, respectively; $\sigma_{1}$ is the measurement
error of $W^{\rm obs}$ as given in Table\,\ref{table:EWs_all}, 
and $\sigma_{2}$ is the observational error due to insufficient
time averaging given the limited field of view covered by the slit
on the solar surface. The error $\sigma_{2}$ is only considered  in case of the
WCLC spectra, where the exposure times are less than the typical granule
lifetime and the slit covers roughly three times the surface area 
of the 3D model. We derive $\sigma_{2}$ from a bootstrap-like
test, where a random subsample of three different snapshots is repeatedly 
drawn from a larger sample of 20 3D snapshots without replacement. 
For each $\mu$-value, we generated $10\,000$ realizations of the
possible appearance of the solar surface in the field of view, using the
same \logAO\ -- \SH\ combination as for the  equivalent width measurement
(see above). The trial-to-trial variance of the derived distribution of
the subsample averaged EWs is then taken as $\sigma_{2}$.
Pereira's spectra have much better statistics (large number of observations
per $\mu$-angle), and there is no need to consider this kind of measurement
error in the case of these spectra. 

For the purpose of EW fitting, we compute a grid of equivalent widths, 
$W$(\SH, \logAO; $\mu$) from the corresponding sets of 3D non-LTE line
profiles, where $0 \le \SH \le 8/3$, $8.65 \le \logAO \le 8.83$, and
$\mu$ covering the observed limb angles. This allows us to compute
$\chi^{2}(\SH,\logAO)$ in the \SH\ -- \logAO\ plane, separately for each 
of the three triplet components, as shown in Fig.\,\ref{fig:EWcontours}
for both observational data sets. The best-fit solution is defined by the
location of the minimum of \chisq, as indicated in the plots by a white plus
sign, which is found by EW interpolation using third-order polynomials.

The results of the EW fitting for various model atmospheres are summarized 
in Table\,\ref{table:EWfit}, listing \logAO\ and \SH\ of the best fit for
the WCLC and Pereira observed spectra. The formal fitting errors of \logAO\ 
and \SH\ are given by the projection of the contour line defined by 
$\chisq - \chi^2_{\rm min} = 1$ (white ellipse in Fig.~\ref{fig:EWcontours}) 
on the \logAO\ and \SH\ axis, respectively. We also provide the reduced \chisq\ 
of the best fit, which is defined as $\chi_{\rm red}^2=\chisq/(m - p$), where 
$p = 2$ is the number of free fitting parameters, as a measure of the quality
of the fit. 

The high quality of the EW fit achieved with the 3D model is also
illustrated in Fig.\,\ref{fig:ctlEWs}, where we show the center-to-limb 
variation of observed and theoretical EWs for the best-fitting parameters 
and for the two extreme values \SH=0 and 8/3. Corresponding plots for all 
1D model atmospheres investigated in this work are provided in 
Appendix\,\ref{app:1DEWs}.
      
To derive the best estimate the solar oxygen abundance from EW fitting, 
we rely on the results of the 3D model atmosphere, which show the smallest
line-to-line scatter and the lowest \chisq. Computing the mean 
and errors for the two data sets as before in Sect.\,\ref{sect:proffit_1},
we obtain 
$\left<{\logAO}\right> = 8.754 \pm 0.010$\,dex (WCLC data set) and 
$\left<{\logAO}\right> = 8.771 \pm 0.006$\,dex (Pereira data set). Again, 
the formal $1\,\sigma$ error bars of these two abundance estimates do not 
overlap (by a narrow margin). Nevertheless, we derive the best estimate by
averaging over the two data sets, arriving at 
$\left<\left<{\logAO}\right>\right> = 8.763 \pm 0.012$, and 
$\left<\left<{\SH}\right>\right> = 1.6 \pm 0.2$. These numbers are fully 
consistent with the final result obtained from the line profile fitting 
described in Section\,\ref{sect:proffit_1}. As in the latter case,
the best-fit solutions, derived from the \xtmean{3D} and 1D \LHD\ modeling,
turn out to be unphysical (negative \SH). We discard these results as  
irrelevant and exclude them from Table\,\ref{table:EWfit} (see also 
Sect.\,\ref{sect:discussion2} below).

\section{Discussion}
\label{sect:discussion}
\subsection{Comparison with the work of Pereira et al.\ (2009)}
\label{sect:discussion1}
In the present work, we simultaneously derived  the oxygen abundance, 
\logAO, and the scaling factor for collisions with neutral hydrogen, \SH,
using two different methods of analyzing the $\mu$-dependent spectra of 
the \ion{O}{i} IR triplet: line profile fitting and equivalent width fitting.
While the former method yields $\logAO = 8.757 \pm 0.010$, 
$\SH  = 1.6 \pm 0.2$, the latter gives $\logAO = 8.763 \pm 0.012$, 
$\SH  = 1.6 \pm 0.2$. In both methods, the spectra at the different 
$\mu$-angles are assigned equal weight.
 
\citet{PAK09} followed a different methodology, using a mix of EW fitting
and line profile fitting, and assigning a distinguished role to the fits
at disk center. More precisely, they first derive \SH\ from simultaneously 
fitting the equivalent widths at all $\mu$-angles and all three triplet 
components. For a given \SH, the oxygen abundance is varied
to find the local minimum \chisq(\SH). Then the minimum of this function 
defines the global minimum of \chisq, which corresponds to their best-fitting 
global value of \SH\,=\,$0.85$. Fixing \SH\ at this value, these authors 
determine the oxygen abundance by fitting the line profiles at disk center, 
separately for each triplet component, 
and average the resulting \logAO\ values to obtain their best estimate of
the oxygen abundance, \mbox{$\logAO = 8.68$.}

In Fig.\,\ref{fig:center_logAO}, we show a comparison of the \logAO\ -- \SH\
relation obtained from our 3D-NLTE modeling with the results of \citet{PAK09},
for each of the three triplet lines and the two different disk-center 
spectra. For a given \SH, the oxygen abundance is determined from 
disk-center line profile fitting, as described in 
Sect.\,\ref{sect:proffit_0} (first sweep, but with fixed \SH).
The figure  immediately shows
that our \logAO\ -- \SH\ relation differs significantly from that of
\citet[][their Table 3]{PAK09}, implying a systematically higher \logAO\
for given \SH. For the value $\SH = 0.85$, which was determined by
\citet{PAK09}, our relation suggests $\logAO \approx 8.71-8.72$. 
This value is clearly incompatible with the value $\logAO \approx 8.64-8.68$,
found by the latter authors. However, we are cautious about making a direct
comparison of the \logAO\ -- \SH\ relations since possible differences in the particular implementation of 
the Drawin formula (definition of \SH) might introduce a bias (see 
Sect.\,\ref{sect:discussion31}). This is the case  with the exception of \SH\,=\,0 and $\infty$ (LTE).

In addition, the line profile fits that we achieve with our 3D non-LTE 
synthetic line profiles (see Fig.\,\ref{fig:ctlProf3D})
are superior to the fitting results shown in the work of \citet{PAK09} 
(see their Fig.~6 and Sect.\,4.1.2.) where, generally, a less perfect 
agreement was found.

We can only speculate about what causes the systematic difference between 
our results and those obtained by \citet{PAK09}.
Differences in the employed oxygen model atom, 
the adopted atomic data (like photoionization and collisional cross sections), 
and other numerical details entering the solution of the non-LTE rate 
equations, are among the prime suspects. Since we even see systematic
\logAO\ differences of about $0.03$~dex in LTE (right column of 
Fig.~\ref{fig:center_logAO}), we conclude that differences in the 3D 
hydrodynamical model atmospheres and / or the spectrum synthesis codes must 
also play a role.

It seems very likely that the above mentioned differences in the theoretical 
3D plus non-LTE modeling are the main reason for the discrepancy in the final
oxygen abundance of almost 0.1~dex between the present work and that of
\citet{PAK09}. The different approaches of analyzing the data constitute
an additional source of uncertainty.

\subsubsection{Impact of equivalent width measurements}
\begin{figure}[!t]
 \centering
 \mbox{\includegraphics[width=9cm]{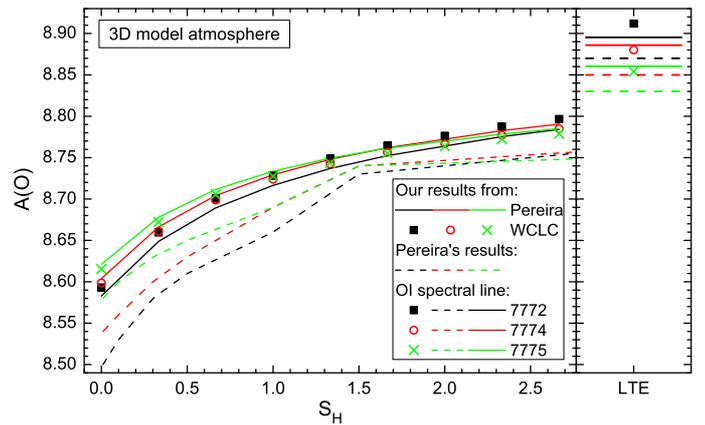}}
 \caption{Oxygen abundance determined from disk-center line profile 
          fitting as a function of \SH\ for the three triplet lines and 
          the two different observed spectra. The 3D-NLTE results of the 
          present work (solid lines, symbols) are compared with those of 
          \citet[][Tab.\,3, dashed lines]{PAK09}, obtained from 
          their profile fits of the same disk-center spectra.}
 \label{fig:center_logAO}
\end{figure}

As described in Sect.\,\ref{sect:EWfit_1}, we systematically derive  
larger equivalent widths than \citet{PAK09} near the disk center and lower 
EWs near the limb, from the same observed spectra (see 
Table\,\ref{table:EWs_all}) due to different measurement methods.
One might therefore ask how much of the difference between the oxygen 
abundance derived in the present work and by Pereira et al.\ can be ascribed 
to differences in the measured equivalent widths. To answer this question, 
we have repeated our 3D, non-LTE, EW fitting procedure with the original 
equivalent widths as given by \citet{PAK09}. We find that the derived oxygen
abundance decreases by $\Delta\logAO\approx -0.045$~dex, explaining roughly
half the discussed \logAO\ discrepancy. We point out, however, that the
almost perfect consistency between profile fitting and EW fitting results
is destroyed if we were to adopt the equivalent widths measured by 
\citet{PAK09}. Note also that the mismatch illustrated in 
Fig.\,\ref{fig:center_logAO} is completely independent of any EW 
measurements.

\begin{figure}[!h]
\mbox{\includegraphics[width=9cm, clip=true]{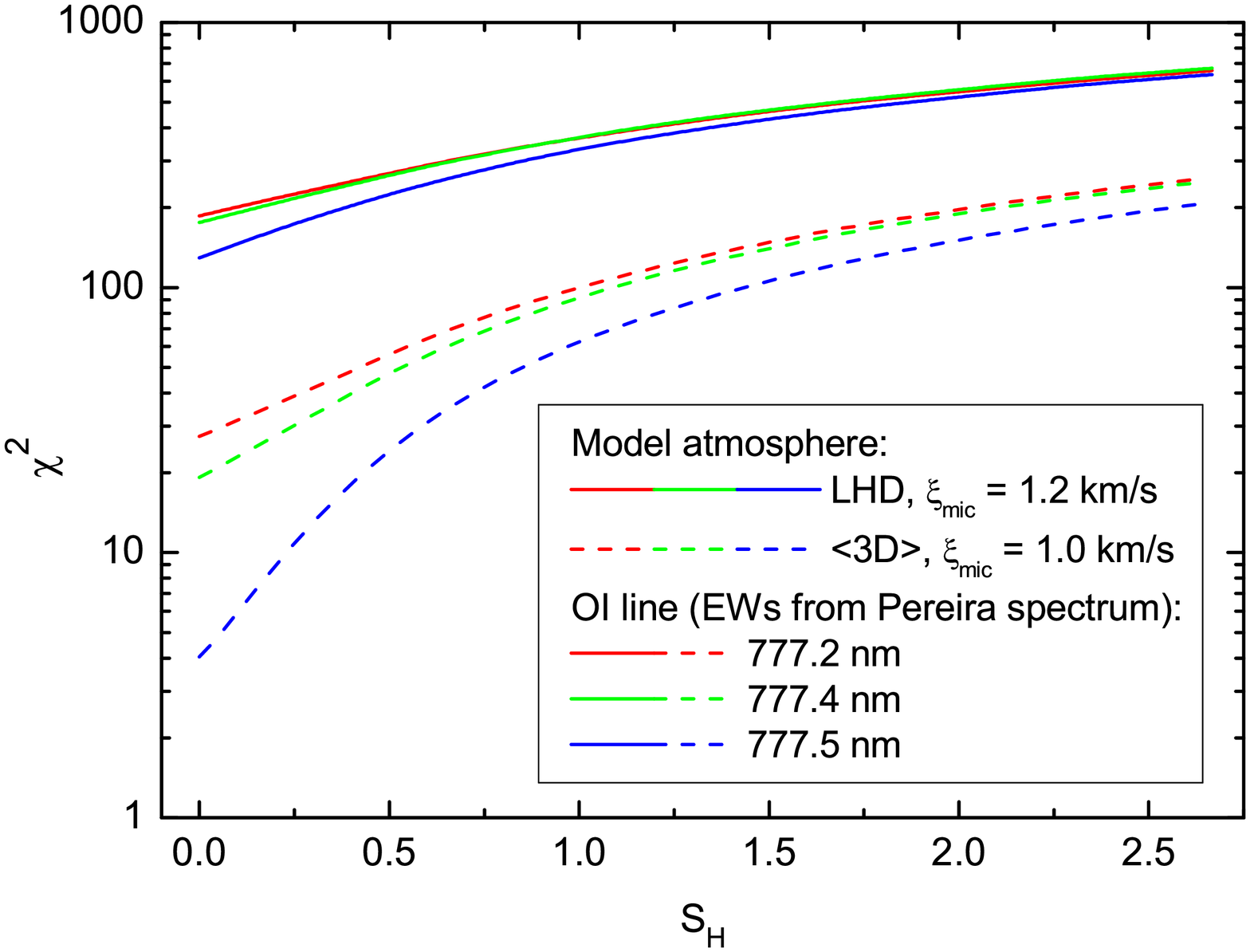}}
\caption{Minimum \chisq\ as a function of \SH\ obtained from 
global EW fitting with the \xtmean{3D}
and \LHD\ model atmospheres. For each of the three triplet lines, we plot 
the minimum of \chisq\ (nonreduced) at fixed \SH. No minimum of \chisq\ 
is found in the range $0 \le \SH \le 8/3$. At \SH=0, the implied oxygen 
abundances are 
\logAO\ = $8.61$, $8.62$, $8.63$ (\xtmean{3D}), and
\logAO\ = $8.59$, $8.60$, $8.61$ (\LHD), for $\lambda\,777.2$, 
$777.4$, and $777.5$\,nm, respectively.
The best fits would be obtained at negative \SH\ and even lower \logAO.}
\label{fig:LHD_EWfit_chi2}
\end{figure}

\subsection{Interpretation of the 1D results}
\label{sect:discussion2}
As mentioned in Sect.\,\ref{sect:proffit}, the oxygen abundance derived from 
line profile fitting based on 1D models (in particular \xtmean{3D}\ and 
\LHD) is significantly lower, by up to $-0.15$~dex, than the value derived
with the 3D modeling. A very similar discrepancy between 1D and 3D results 
is obtained from the method of equivalent width fitting. The main reason is
that the 1D models predict a steeper gradient of the line strength with
respect to limb angle $\mu$ (compare Fig.\,\ref{fig:ctlEWs} with
Figs.\,\ref{fig:ctlEWs_3Dmeanmic10} and \ref{fig:ctlEWs_LHDmic12}). 
Under such circumstances, the best global
fit of the observed center-to-limb variation of the triplet lines would formally
require a negative \SH, as demonstrated in Fig.~\ref{fig:LHD_EWfit_chi2}.
Because of the tight correlation between \SH\ and \logAO\ (see 
Fig.\,\ref{fig:center_logAO}), a low \SH\ implies a low oxygen abundance: 
while $\SH\approx 1.6$ for the 3D model, $\SH\equiv 0$ for the \xtmean{3D} 
and \LHD\ models, explaining the exceedingly low oxygen abundance derived 
with these 1D models. We add that our findings are in line with the results
of \citet[][their Fig.\,5]{PAK09}.

\subsubsection{The microturbulence problem}
Since the triplet lines are partially saturated, their formation is sensitive 
to the choice of the microturbulence parameter, \vmic, to be used with the
1D models. While the value of \vmic\ can be determined empirically, 
the main issue in the present context is the common assumption that 
\vmic\ is independent of $\mu$. In fact, it is well known empirically that 
both micro- and macroturbulence increase towards the solar limb 
\citep[e.g.,][]{Holweger+al78}. Evidently, we could improve the quality
of the 1D fits by introducing a $\mu$-dependent microturbulence parameter.
With a more realistic \vmic\ increasing towards the limb, the slope of 
$W(\mu)$ at fixed \logAO\ would diminish, such that the best fit would 
require a larger (positive) \SH\ and hence a higher oxygen abundance. 
Although it may be worthwhile, we have refrained from exploring the effects 
of a $\mu$-dependent microturbulence on the results of the 1D models in the
present work.  

It is obvious that the 1D models with a fixed microturbulence are unrealistic,
and lead to unphysical conclusions about the scaling factor ($\SH < 0$).
The oxygen abundance derived from these 1D models for the \ion{O}{i} IR 
triplet lines is therefore systematically biased towards exceedingly low values.
Even though the analysis based on the HM model atmosphere
yields positive \SH\ values, the resulting \logAO\ is considered as
misleading, too, since the HM model suffers from the same microturbulence
issues as the \xtmean{3D} and \LHD\ models, albeit in a less obvious way,
because the problems are somehow masked by the empirical temperature 
structure of the HM model.
Consequently, we have ignored all 1D results for the estimation of the 
final solar oxygen abundance. We rather rely on the 
analysis with the 3D model, which has a built-in hydrodynamical velocity 
field that properly represents a $\mu$-dependent micro- and macroturbulence.

\subsubsection{1D -- 3D comparison}
As described above, the microturbulence problem renders a direct comparison
of the abundance determinations from 1D and 3D models meaningless. 
Nevertheless, it is possible to ask what oxygen abundance would be obtained
from the 1D models if we use the same \SH-value as found in the 3D analysis,
i.e., considering this \SH-value as a property of the model atom.
Fixing \SH\ in this way, we  determined \logAO\ from the 1D models by 
fitting the observed equivalent width at disk center, assuming the canonical
microturbulence value of \vmic\ = 1.0\,km/s. The results are summarized in
Table\,\ref{table:EWfitmu1}.

Remarkably, the HM model atmosphere requires almost exactly
the same oxygen abundance as the 3D model to match the measured disk-center
equivalent width. This is indeed what is expected from a semi-empirical model.
The averaged 3D model gives only a slightly lower \logAO, suggesting that the
effect of horizontal inhomogeneities is of minor importance, at least for the
line formation at $\mu=1$. In contrast, the LHD model indicates a
significantly lower oxygen abundance, which we attribute to the somewhat 
steeper temperature gradient in the line forming region 
(see Sect.\,\ref{sect:HMull}).

\begin{table}
\caption{Results of non-LTE EW fitting with various model 
atmospheres for given disk-center equivalent width $W_1$ 
(as measured from the Pereira data set, see Table\,\ref{table:EWs_all}) 
and \SH-factor (as obtained from the global EW fit with the 3D model, 
see Table\,\ref{table:EWfit}). The oxygen abundance \logAO\ required  
to reproduce $W_1$ with the 3D, averaged 3D, HM, and LHD model, respectively, 
is listed in Cols.\,(4) -- (7). For all 1D models, a microturbulence of 
\vmic\ = 1.0\,km/s is assumed.}
\label{table:EWfitmu1}
\centering
\begin{tabular}{|c|c c||c c c c|}
\hline\noalign{\smallskip}
                $\lambda$ & $W_1$ & \SH\ & \multicolumn{4}{|c|}{\logAO} \\
                {[nm]}    & [pm]  &      & 3D & \xtmean{3D} & HM & LHD  \\
\hline\noalign{\smallskip}
777.2 &  8.63 & 1.67 & 8.758 & 8.744 & 8.753 & 8.686 \\
777.4 &  7.55 & 1.69 & 8.766 & 8.753 & 8.765 & 8.695 \\
777.5 &  5.99 & 1.96 & 8.770 & 8.759 & 8.778 & 8.703 \\
Mean  & --    & --   & 8.765 & 8.752 & 8.765 & 8.695 \\
\hline
\end{tabular} 
\end{table}

\subsection{Treatment of inelastic collisions by neutral hydrogen}
\label{sect:discussion3}

Collisions of neutral oxygen atoms with neutral hydrogen give rise to
bound-bound (excitation) and bound-free (ionization) transitions.  As shown
previously, they play a very important role in establishing the statistical
equilibrium level population of \ion{O}{i}. Unfortunately, quantum mechanical
calculations or experimental measurements of the relevant collisional cross
sections are so far unavailable. We have therefore resorted to calculating the
excitation and ionization rates due to collisions by neutral hydrogen by the
classical Drawin formula. Following common
practice, we  scaled the resulting rates by an adjustable factor, \SH,
which is determined from the best fit to the observations together with the
oxygen abundance, as detailed above. 

\subsubsection{Alternative implementations of the Drawin formula}
\label{sect:discussion31}
In the present work, the cross sections for inelastic collisions
with neutral hydrogen are computed according to the classical recipe of 
Drawin \citep{DRAW69, SH84, Lam93}, and hence are proportional to the 
oscillator strength $f$ for all bound-bound transitions. For optically 
forbidden transitions, the collisional cross section is therefore zero.
All hydrogen related collision rates included in our oxygen model atom, 
bound-bound and bound-free, are then scaled by the same factor \SH.
It is not clear whether other authors \citep[e.g.,][]{PAK09} might
have adopted different assumptions. We have therefore investigated two test
cases to understand the consequences for the derived oxygen abundance of 
(i) allowing for collisional coupling through optically forbidden transitions, 
and (ii) assuming a constant collision strength independent of $f$ for all
allowed transitions.

To reduce the computational cost, we have restricted the test
calculations to EW fitting of the Pereira data set, using the HM 1D model 
atmosphere with \vmic=1.2\,km/s. The results are summarized in 
Table\,\ref{table:Drawintest}. The oxygen abundance and \SH-factor 
obtained with the standard setup are shown in Cols.\,(2) and (3) (`Case 0',
cf.\, Table\,\ref{table:EWfit}).

In `Case 1', we have included collisional
excitation for all possible transitions according to the Drawin formula,
where the cross sections of the optically allowed transitions are proportional
to the oscillator strength $f$ (as in `Case 0'). In this case, an effective 
minimum oscillator strength of $f$\,=\,$0.001$ is adopted for all forbidden
transitions, including the transitions between the triplet and quintuplet 
system, as suggested by \citet{FAB09}. 
Scaling by \SH\ is restricted to the allowed transitions. The effect of
the additional collisional coupling via the forbidden transitions on the
oxygen abundance derived from EW fitting of the triplet lines is marginal:
on average, \logAO\ is reduced by only $0.004$~dex.

In `Case 2', the treatment of the collisional coupling in the
forbidden transitions is the same as in `Case 1', but the cross sections 
of all optically allowed transitions are computed with $f$\,=\,$1$.
Again, scaling by \SH\ is restricted to the allowed transitions. The effect 
of removing the proportionality between collision strength and oscillator
strength for the allowed transitions on the derived oxygen abundance amounts
to $\approx -0.01$~dex, while the numerical value of the required \SH\
is reduced by roughly a factor 2 with respect to `Case 0'.

The effects shown in Table\,\ref{table:Drawintest} are expected
to be of similar magnitude in the full 3D analysis. We conclude from this 
experiment that the oxygen abundance derived from the non-LTE analysis
of the \ion{O}{i} IR triplet lines is not particularly sensitive to the
details of the implementation of the Drawin formula, even if the numerical
value of \SH\ can differ significantly.

\begin{table}
\caption{Resulting oxygen abundance \logAO\ and scaling factor \SH\
derived from non-LTE EW fitting of the Pereira data set,
using the HM model atmosphere with \vmic=1.2\,km/s. Cases (0) -- (2)
refer to different assumptions about the excitation rates due to 
collisions by neutral hydrogen (see text for details).}
\label{table:Drawintest}
\centering
\begin{tabular}{|c|c c | c c | c c|}
\hline\noalign{\smallskip}
                $\lambda$ & \multicolumn{2}{|c|}{Case 0} &
                            \multicolumn{2}{|c|}{Case 1} &
                            \multicolumn{2}{|c|}{Case 2} \\
                {[nm]}    & \logAO & $\SH^{(0)}$ & 
                          $\Delta\logAO$ & $\SH^{(1)}/\SH^{(0)}$ & 
                          $\Delta\logAO$ & $\SH^{(2)}/\SH^{(0)}$ \\
\hline\noalign{\smallskip}
$777.2$ &  $8.650$ & 0.36 & $-0.004$ & 0.818 & $-0.011$ & 0.515 \\
$777.4$ &  $8.674$ & 0.42 & $-0.005$ & 0.854 & $-0.014$ & 0.512 \\
$777.5$ &  $8.721$ & 0.80 & $-0.003$ & 1.000 & $-0.006$ & 0.722 \\
Mean    &  $8.682$ & 0.52 & $-0.004$ & 0.891 & $-0.010$ & 0.583 \\
\hline
\end{tabular} 
\end{table}

\subsubsection{Assumption of a universal \SH}
\label{sect:discussion32}
Although this approach is commonly accepted,  the assumption of a
universal scaling factor is questionable. It could well be that each line is
characterized by a slightly different value of \SH, and / or that individual
bound-bound and bound-free transitions require significantly different scaling
factors. This would not be surprising, as it is well known that the classical 
treatment of collisions does not have a rigorous physical foundation. In fact,
such nonuniform scaling has recently been demonstrated for Mg $+$ H by
\citet{Bark12}. In principle, \SH\ should depend on temperature
\citep[see][]{BBG11} and, because the different triplet lines form at
different atmospheric heights, \SH\ could be expected 
to vary slightly from line to line. Our analysis allows for this 
possibility, since we treat each of the triplet components separately. 
However, we do not find any systematic trend in \SH\ with
wavelength, and consider the line-to-line variation of \SH\ as random,
reflecting the intrinsic uncertainties of our method. 
With this reasoning, a systematic variation of \SH\ as a function of 
the limb angle also seems  unlikely, and our analysis assumes a 
$\mu$-independent \SH. In summary, the treatment of the collisions 
by neutral hydrogen in terms of the empirical scaling factor \SH\ appears
unsatisfactory.

\section{Conclusions}
\label{sect:conclusions}

The present analysis of the \ion{O}{i} IR triplet is based on two data sets 
of intensity spectra recorded at different $\mu$-angles across the solar disk: 
a set of spectrograms taken by one of the authors (WCL) with the rapid scan 
double-pass spectrometer at the McMath-Pierce Solar Telescope, Kitt Peak, 
on 12 September 2006, and the set of high-quality spectra used by 
\citet{PAK09}, observed with the Swedish 1-m Solar Telescope on Roque de 
Los Muchachos, La Palma, in May 2007. 

The photospheric oxygen abundance, \logAO, and the scaling factor for
collisional excitation / ionization by neutral hydrogen, \SH, were determined
simultaneously, using two different \chisq\ fitting methods of minimizing the
difference between observed and synthetic spectra. In the first method, we 
simultaneously fit the \emph{line profiles} at all $\mu$-angles. Averaging
the results of the three triplet components and the two different sets of 
observed spectra, we find an oxygen abundance of 
\mbox{$\logAO = 8.757 \pm 0.010$} \mbox{($\SH = 1.6 \pm 0.2$)}, where 
the error is defined by the standard deviation 
of the individual results. In the second method, we fit \emph{equivalent
widths} instead of line profiles, and obtain a perfectly consistent result,
$\logAO = 8.763 \pm 0.012$ ($\SH = 1.6 \pm 0.2$).

The above numbers were derived by employing 3D non-LTE synthetic spectra, 
consistently taking  non-LTE line formation in a 3D hydrodynamical
model of the solar atmosphere into account. When performing the same analysis 
with synthetic spectra computed from various 1D model atmospheres, we obtain
much less perfect line profile fits, a larger line-to-line variation of
\logAO\ and \SH, and a significantly lower value of the best-fitting oxygen 
abundance. We argue that the 1D results are systematically biased
towards low \SH\ and \logAO\ values, mainly as a consequence of the assumption
of a $\mu$-independent microturbulence.  Consequently, we have ignored all 1D
results for the estimation of the final solar oxygen abundance in the present
work. We rather rely on the analysis with the 3D model, which has a built-in
hydrodynamical velocity field that properly represents a $\mu$-dependent
micro- and macroturbulence.

All individual results derived from the various observations of the \ion{O}{i}
IR triplet lines with different analysis methods, fall in the range
$8.74 < \logAO < 8.78$. We thus arrive at a final 3D non-LTE oxygen abundance 
of $\logAO = 8.76 \pm 0.02$. This coincides with the value recommended in
Paper~I \citep[see also][]{CLS11} based on various atomic \ion{O}{i} lines,
but is significantly higher than suggested by the analysis of \citet{PAK09}, 
who find $\logAO = 8.68$. The reason for this discrepancy of almost $0.1$\,dex 
is unknown. Presumably, differences in the 3D hydrodynamical model atmospheres 
(\COBOLD\ versus {\tt STAGGER}), in the employed oxygen model atom, or in the 
adopted atomic data may offer an explanation.

The present detailed investigation leads us to conclude that,
from the \COBOLD\ perspective, a solar photospheric oxygen abundance 
lower than \mbox{\logAO\ = 8.70} now appears very unlikely. In view of 
the recent laboratory measurements of \citet{Bailey2015}, indicating that 
the iron opacity near the solar \mbox{radiation /} convection zone boundary 
is significantly higher than assumed so far, the conflict between surface 
chemical abundances and helioseismic models might thus be less severe than 
previously thought.

Finally, we note that the treatment of the collisions by neutral hydrogen
in terms of an empirical scaling factor \SH\ is far from satisfactory. 
The assumption that the same \SH\ is valid for all transitions is 
questionable and introduces an additional systematic uncertainty of 
unknown magnitude in the determination of the photospheric solar oxygen 
abundance from the \ion{O}{i} IR triplet. Clearly, accurate theoretical 
or experimental collisional cross sections for neutral oxygen are badly needed.


\begin{acknowledgements}
{Part of this work was supported by grants from the Research Council of
Lithuania (MIP-065/2013) and the French-Lithuanian programme ``Gilibert'' (TAP
LZ 06/2013, Research Council of Lithuania). MS and HGL acknowledge the funding
of research visits to Vilnius by the Research Council of Lithuania. 
DP acknowledges financial support from the German Academic Exchange 
Service (DAAD)
and AIP (Potsdam), which funded visits to Potsdam. EC is grateful to the 
FONDATION MERAC for funding her fellowship. HGL acknowledges financial 
support by the Sonderforschungsbereich SFB881 ``The Milky Way System'' 
(subproject A4) of the German Research Foundation (DFG). The authors
are also grateful to Sergii Korotin, Lyudmila Mashonkina, 
Tatyana Sitnova, and Paul Barklem for discussions regarding the 
non-LTE treatment of oxygen spectral lines. We thank the referee, 
Remo Collet, for his competent and constructive suggestions that
helped us to further improve the presentation of our results.}
\end{acknowledgements}


\bibliographystyle{aa}

\begin{appendix}

\section{The oxygen model atom designed for solar applications}
\label{sect:app-OIatom}

In Tables \ref{tab:oxygen-levels} and \ref{tab:oxygen-trans}, we provide 
details about the $22$ energy levels and $54$ radiative transitions 
considered in the oxygen model atom used in this work. The data are 
taken from the NIST Atomic Spectra Database 
(\url{http://www.nist.gov/pml/data/asd.cfm/}).

\begin{table}[!h]
\caption{Atomic energy levels of the oxygen model atom used in this work.}
\label{tab:oxygen-levels}
\centering
\begin{tabular}{c c c c c}
\hline\noalign{\smallskip}
(1)   & (2)                & (3)   & (4)                  & (5)          \\
Level & Configuration      & \multicolumn{2}{c}{Energy} & Statistical  \\
\#    & and term           & [Ryd] &  [eV]                &   weight     \\
\hline\noalign{\smallskip}
 1    & 2p~$^{3}$P          & 0.0000000 & 0.0000 &   9  \\
 2    & 2p~$^{1}$D          & 0.1438880 & 1.9577 &   5  \\
 3    & 2p~$^{1}$S          & 0.3079407 & 4.1898 &   1  \\
 4    & 3s~$^{5}$S$^{o}$    & 0.6722253 & 9.1461 &   5  \\
 5    & 3s~$^{3}$S$^{o}$    & 0.6990967 & 9.5117 &   3  \\
 6    & 3p~$^{5}$P          & 0.7894125 & 10.741 &  15  \\
 7    & 3p~$^{3}$P          & 0.8069543 & 10.979 &   9  \\
 8    & 4s~$^{5}$S$^{o}$    & 0.8700480 & 11.838 &   5  \\
 9    & 4s~$^{3}$S$^{o}$    & 0.8768672 & 11.930 &   3  \\
 10   & 3d~$^{5}$D$^{o}$    & 0.8877632 & 12.079 &  25  \\
 11   & 3d~$^{3}$D$^{o}$    & 0.8883803 & 12.087 &  15  \\
 12   & 4p~$^{5}$P          & 0.9030074 & 12.286 &  15  \\
 13   & 4p~$^{3}$P          & 0.9076500 & 12.349 &   9  \\
 14   & 3s~$^{3}$D$^{o}$    & 0.9216869 & 12.540 &  15  \\
 15   & 5s~$^{5}$S$^{o}$    & 0.9305559 & 12.661 &   5  \\
 16   & 5s~$^{3}$S$^{o}$    & 0.9332467 & 12.698 &   3  \\
 17   & 4d~$^{5}$D$^{o}$    & 0.9373803 & 12.754 &  25  \\
 18   & 4d~$^{3}$D$^{o}$    & 0.9377701 & 12.759 &  15  \\
 19   & 4f~$^{5}$F          & 0.9383157 & 12.766 &  35  \\
 20   & 4f~$^{3}$F          & 0.9383165 & 12.766 &  21  \\
 21   & 5p~$^{5}$P          & 0.9443120 & 12.848 &  15  \\
 22   & 5p~$^{3}$P          & 0.9465332 & 12.878 &   9  \\
\hline
\end{tabular}
\end{table}

\begin{table}[!h]
\caption{Radiative transitions included in the oxygen model atom 
used in this work.}
\label{tab:oxygen-trans}
\centering

\begin{tabular}{c r r r r r r}

\hline\noalign{\smallskip}
1)        & (2)       & (3)               & (4)        &        (5)         &  (6)      & (7)                                \\
Line & \multicolumn{2}{c}{Levels}  & \multicolumn{1}{c}{$\lambda$} & Oscillator         & \multicolumn{2}{c}{Transition probability}   \\
\#         & lo   & up             & [$\AA$]      & strength, $f$  &  $A_{ji}$   & $B_{ij}$                             \\
\hline\noalign{\smallskip}
  1        &  1        &  2                &    6372.0  &   3.57E-11     & 1.07E-2     & 3.80E+00                             \\
  2        &  1        &  5                &    1303.5  &   5.19E-02     & 6.12E+8     & 1.13E+09                             \\
  3        &  1        &  9                &    1040.1  &   9.17E-03     & 1.70E+8     & 1.60E+08                             \\
  4        &  1        & 10                &    1026.5  &   8.33E-13     & 1.90E-3     & 1.43E-02                             \\
  5        &  1        & 11                &    1026.6  &   2.00E-02     & 7.63E+7     & 3.45E+08                             \\
  6        &  1        & 16                &     977.2  &   3.30E-03     & 6.93E+7     & 5.41E+07                             \\
  7        &  1        & 17                &     974.1  &   5.62E-05     & 1.43E+5     & 9.18E+05                             \\
  8        &  1        & 18                &     972.5  &   1.37E-02     & 5.84E+7     & 2.24E+08                             \\
  9        &  2        &  3                &    5578.9  &   1.16E-09     & 1.26E+0     & 1.08E+02                             \\
 10        &  2        &  4                &    1727.1  &   7.76E-10     & 1.74E+0     & 2.24E+01                             \\
 11        &  2        &  5                &    1641.3  &   4.43E-07     & 1.83E+3     & 1.22E+04                             \\
 12        &  2        & 14                &    1172.6  &   8.51E-06     & 1.38E+4     & 1.67E+05                             \\
 13        &  3        &  5                &    2324.8  &   1.12E-08     & 4.61E+0     & 4.40E+02                             \\
 14        &  3        & 14                &    1484.5  &   8.97E-06     & 1.81E+3     & 2.23E+05                             \\
 15        &  4        &  6                &    7775.5  &   1.00E+00     & 3.69E+7     & 1.31E+11                             \\
 16        &  4        & 12                &    3948.5  &   3.42E-03     & 4.89E+5     & 2.27E+08                             \\
 17        &  4        & 21                &    3348.2  &   3.02E-04     & 5.99E+4     & 1.69E+07                             \\
 18        &  5        &  7                &    8448.8  &   1.03E+00     & 3.22E+7     & 1.46E+11                             \\
 19        &  5        & 13                &    4369.5  &   6.50E-03     & 7.58E+5     & 4.77E+08                             \\
 20        &  5        & 22                &    3692.4  &   5.67E-04     & 9.30E+4     & 3.50E+07                             \\
 21        &  6        &  8                &   11302.5  &   1.70E-01     & 2.67E+7     & 3.23E+10                             \\
 22        &  6        & 10                &    9266.4  &   9.54E-01     & 4.45E+7     & 1.48E+11                             \\
 23        &  6        & 15                &    6456.8  &   1.71E-02     & 8.25E+6     & 1.86E+09                             \\
 24        &  6        & 17                &    6159.0  &   7.22E-02     & 7.62E+6     & 7.46E+09                             \\
 25        &  7        &  9                &   13168.2  &   1.81E-01     & 2.14E+7     & 3.97E+10                             \\
 26        &  7        & 11                &   11289.9  &   9.67E-01     & 3.09E+7     & 1.81E+11                             \\
 27        &  7        & 14                &    7992.0  &   8.84E-04     & 5.61E+4     & 1.17E+08                             \\
 28        &  7        & 16                &    7256.4  &   1.74E-02     & 6.72E+6     & 2.11E+09                             \\
 29        &  7        & 18                &    7004.1  &   4.28E-02     & 3.53E+6     & 5.00E+09                             \\
 30        &  8        & 12                &   27644.6  &   1.47E+00     & 4.29E+6     & 6.84E+11                             \\
 31        &  8        & 21                &   12267.7  &   1.67E-02     & 2.47E+5     & 3.44E+09                             \\
 32        &  9        & 13                &   28935.2  &   1.61E+00     & 4.10E+6     & 8.03E+11                             \\
 33        &  9        & 22                &   13076.9  &   2.17E-02     & 2.83E+5     & 4.78E+09                             \\
 34        & 10        & 12                &   59762.0  &   1.59E-01     & 4.97E+5     & 1.60E+11                             \\
 35        & 10        & 19                &   18026.1  &   1.00E+00     & 1.48E+7     & 3.05E+11                             \\
 36        & 10        & 20                &   18021.2  &   1.96E-02     & 4.79E+5     & 5.93E+09                             \\
 37        & 10        & 21                &   16114.7  &   9.43E-04     & 4.04E+4     & 2.55E+08                             \\
 38        & 11        & 13                &   45608.5  &   1.98E-01     & 9.86E+5     & 1.57E+11                             \\
 39        & 11        & 19                &   18244.1  &   2.98E-02     & 2.56E+5     & 9.13E+09                             \\
 40        & 11        & 20                &   18248.6  &   1.02E+00     & 1.47E+7     & 3.14E+11                             \\
 41        & 11        & 22                &   15670.2  &   4.04E-03     & 1.83E+5     & 1.06E+09                             \\
 42        & 12        & 15                &   33083.7  &   3.01E-01     & 5.51E+6     & 1.67E+11                             \\
 43        & 12        & 17                &   26514.7  &   1.13E+00     & 6.45E+6     & 5.04E+11                             \\
 44        & 13        & 14                &   68287.3  &   1.43E-02     & 1.36E+4     & 1.56E+10                             \\
 45        & 13        & 16                &   36617.4  &   3.05E-01     & 4.82E+6     & 1.82E+11                             \\
 46        & 13        & 18                &   30985.0  &   1.15E+00     & 5.04E+6     & 5.85E+11                             \\
 47        & 14        & 20                &   54797.7  &   9.83E-03     & 1.56E+4     & 9.04E+09                             \\
 48        & 14        & 22                &   36687.2  &   1.98E-03     & 1.64E+4     & 1.22E+09                             \\
 49        & 15        & 21                &   66237.7  &   1.88E+00     & 9.57E+5     & 2.10E+12                             \\
 50        & 16        & 22                &   68568.3  &   1.94E+00     & 9.17E+5     & 2.23E+12                             \\
 51        & 17        & 19                &  974107.5  &   4.78E-02     & 2.40E+2     & 7.81E+11                             \\
 52        & 17        & 21                &  131456.0  &   3.14E-01     & 2.02E+5     & 6.93E+11                             \\
 53        & 18        & 20                & 1668808.0  &   2.73E-02     & 4.68E+1     & 7.64E+11                             \\
 54        & 18        & 22                &  103964.9  &   3.53E-01     & 3.63E+5     & 6.16E+11                             \\
\hline
\end{tabular}
\end{table}

\section{Ignoring the fine-structure splitting of the triplet}
\label{app:finesplit}

In Sect.\,\ref{sect:modelatom}, we  mentioned that the upper level 
of the \ion{O}{i} IR triplet (3p$^{5}$P) is treated as a single superlevel 
in our oxygen model atom, thus ignoring the fine structure splitting of this
term. Since the energy differences between fine-structure levels are
negligible, the LTE population of this superlevel equals the sum of those of
the individual sublevels. If the fine-structure levels are merged into one 
superlevel, the three transitions making up the triplet are merged into a
single combined transition, and the $gf$-value of this combined transition
is the sum of the $gf$-values of the individual fine-structure components.

The radiative transition probability (Einstein coefficient or, equivalently,
$gf$-value) of a bound-bound transition plays a two-fold role in the non-LTE
problem: firstly, the photoexcitation rate at given radiation field scales
with $gf$; and secondly, the line opacity is proportional to $gf$, and hence
the radiation field of the transition is a (non-linear) function of the
$gf$-value. Since the three triplet components are well separated in
wavelength and partially saturated in the Sun, merging them leads to a
strongly saturated single line where the mean radiation field in the line,
$\bar{J}$, is diminished below the value in any of the individual original
triplet components. The photoexcitation rates in the combined transition
would therefore be significantly underestimated.

To avoid this systematic bias, we reduce the line opacity of the 
combined triplet transition by a factor 3. This is equivalent to using, 
only for the calculation of $\bar{J}$, the $gf$-value of the intermediate 
triplet component instead of the sum of the $gf$-values of all components. 
For the calculation of the photoexcitation rates at given radiation field, 
however, the total $gf$-value (i.e.,\ the sum of the $gf$-values of the
individual fine structure components) is applied. This procedure guarantees
both $\bar{J}$ and photoexcitation rates to be consistent with
the real situation. 

In the following, we show that the adopted approach is equivalent to having a
detailed multiplet transition with three lines, with highly efficient
collisions between the fine-structure levels. The latter approach was utilized
in numerous works \citep[e.g.,][]{Kis93, PBB00, AGS04} and is assumed to be a
valid representation of the real situation. To demonstrate the
equivalence of the two methods, we  constructed a model atom that includes
the fine-structure splitting of the 3p$^{5}$P level. We  used a solar
\LHD\ model atmosphere to compute the departure coefficients for three cases:

\begin{enumerate}
\item Using the standard model atom where 3p$^{5}$P is treated as a 
      superlevel and $gf$ is reduced by 1/3 for the calculation of the 
      line opacity;
\item Using a model atom that includes the splitting of the 
      3p$^{5}$P term and no collisional coupling between the fine-structure 
      levels;
\item Using a model atom that includes the splitting of the 
      3p$^{5}$P term and strong collisional coupling between the 
      fine-structure levels.
\end{enumerate}
   
\begin{figure}[!t]
 \centering
 \includegraphics[width=9cm]{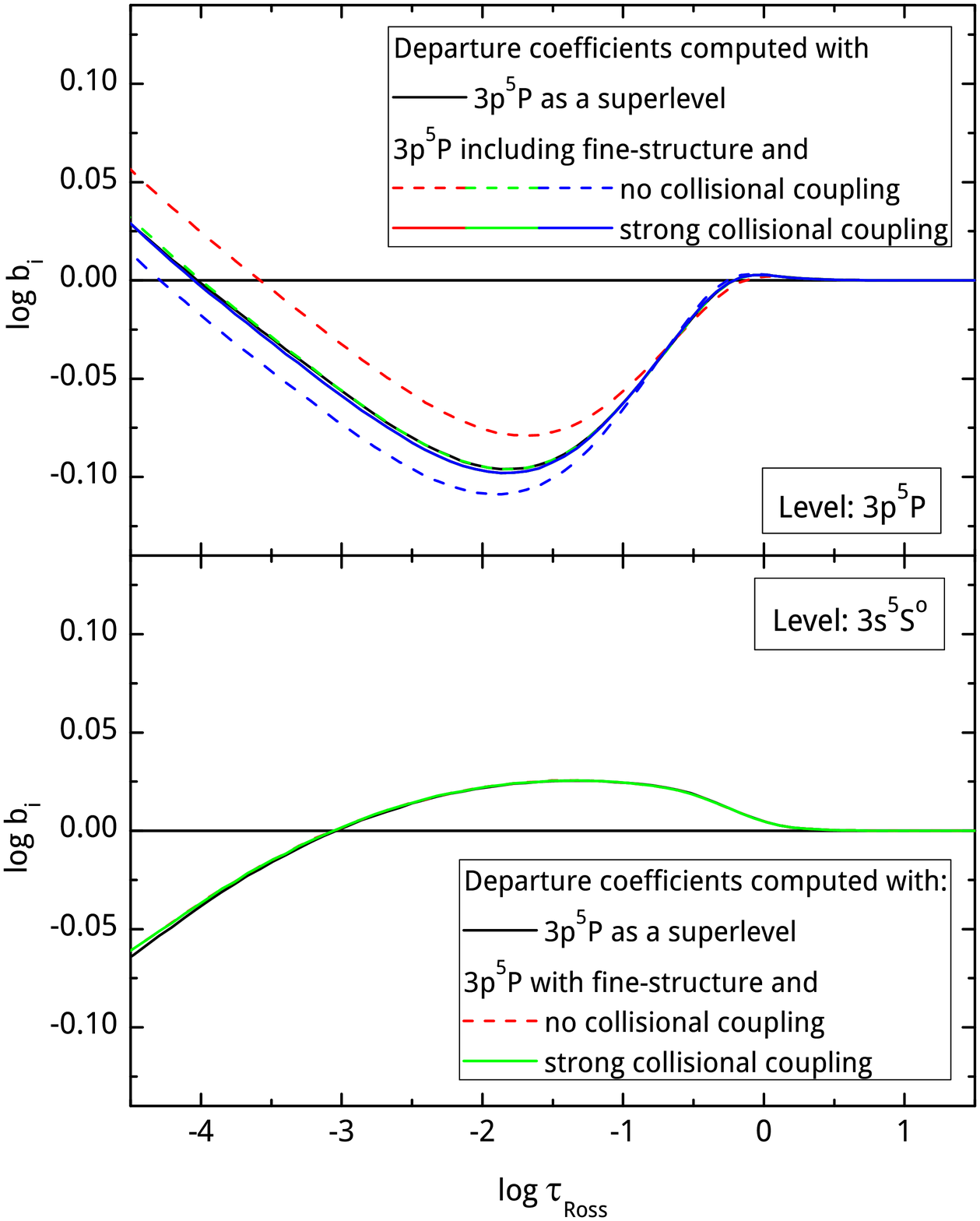}
 \caption{
Departure coefficients of the \ion{O}{i} IR triplet levels as a function of
the optical depth for three cases of the fine-structure treatment (see text
for details). The black solid line shows the results obtained when 3p$^{5}$P 
is treated as a superlevel (case~1); colored lines show the results when
fine-structure splitting is included; dashed lines show case~2 (no collisional 
coupling between fine-structure levels);  solid lines show case~3 (strong 
collisional coupling between fine-structure levels). Top and bottom
panels refer to terms 3p$^{5}$P (upper level of the triplet) and 
3s$^{5}$S$^{o}$ (lower level of the triplet), respectively.}
 \label{fig:finestruct}
\end{figure}

In the last two cases, no scaling of the line opacity is applied. Except for
the treatment of the  3p$^{5}$P term and the transitions that connect to it, 
all three model atoms are identical
(see Sect.\,\ref{sect:modelatom}). For the test calculations, the oxygen 
abundance was set to $\logAO = 8.77$ and $\SH = 1.0$.
   
The results of this experiment are shown in Fig.~\ref{fig:finestruct}. 
Firstly, it is clear that the treatment of the level 3p$^{5}$P does not 
significantly influence the lower level population of the triplet. This 
means that non-LTE line opacity of the \ion{O}{i} IR triplet is not sensitive 
to the details regarding the fine-structure of 3p$^{5}$P. Secondly, it is evident 
that, when no collisional coupling is applied between fine-structure levels, 
they exhibit slightly different departure coefficients. Our analysis
(not detailed here) has shown that this is caused by the different line
strengths of the three triplet components, which leads to different 
photoexcitation rates in each of the triplet transitions and, in turn,
to different departure coefficients. Finally, the most important result is 
that, when efficient collisions between the fine-structure levels are applied, 
the departure coefficients of the individual sublevels thermalize to the 
same values that were computed with the superlevel approach. We consider 
this as a validation of our simplified approach, which reduces the number 
of lines for which the radiation transfer has to be computed and thus facilitates a faster solution (by about $60$\,\%) of the non-LTE statistical equilibrium 
problem.

\section{Effect of horizontal resolution on the 3D-NLTE synthetic spectrum}
\label{app:skipping}

In Sections~\ref{sect:nlte} and~\ref{sect:specsyn}, we have mentioned that to speed up the calculations, we  applied a horizontal reduction 
of the model atmosphere, using only every third vertical column in each
horizontal direction (i.e., discarding 90\% of the data defining the 3D
model atmosphere). In this section, we briefly validate this approach.
   
\begin{figure}[!t]
 \centering
 \includegraphics[width=9cm]{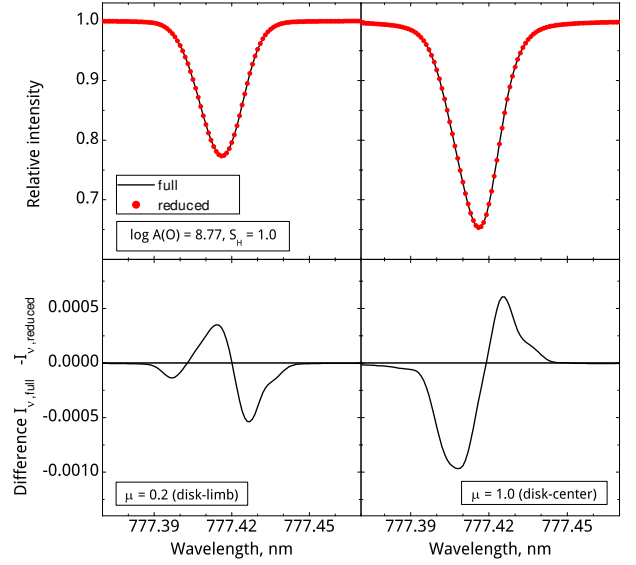}
 \caption{
Effect of reduced horizontal resolution on the 3D-NLTE spectral line profile 
of the middle component of the \ion{O}{i} IR triplet. {\it Top panels}: 
synthetic spectral lines computed with full resolution (black line) and with 
resolution reduced by a factor 3 in each horizontal direction (red
dots). {\it Bottom panels}: difference in residual intensity between 
synthetic spectral lines computed with full and reduced horizontal resolution.
A negative difference means that the spectral line computed with full
resolution is stronger at the particular wavelength. Left and right panels 
show the results at $\mu = 0.2$ and $\mu = 1.0$, respectively.}
 \label{fig:skipping}
\end{figure}

To estimate the effect of the reduced horizontal resolution, we 
 performed non-LTE calculations (Sect.\,\ref{sect:nlte}) and subsequent 
spectrum synthesis (Sect.\,\ref{sect:specsyn}) with a fully resolved single 
3D model snapshot. In these calculations, we  set $\logAO = 8.77$ and 
$\SH = 1.0$. To see how the effect changes across the solar disk, 
spectrum synthesis was performed at two inclination angles: $\mu = 1.0$
and $\mu=0.2$.
   
Fig.~\ref{fig:skipping} shows the results of this experiment. It is seen
that the effect of the reduced horizontal resolution does not exceed 0.1\%
of the continuum intensity anywhere in the line profile. The effect on 
equivalent widths is even smaller, $\delta\,W / W \approx 7\times 10^{-4}$ 
and $12\times 10^{-4}$ for $\mu=0.2$ and $\mu=1.0$, respectively. 
Hence we may conclude that the reduced horizontal resolution that we have 
imposed on the spectrum synthesis does not influence the results in any 
significant way.

\section{Influence of the mixing-length parameter on 1D-NLTE spectrum 
         synthesis of the \ion{O}{i} IR triplet lines}
\label{app:mlp}

The \ion{O}{i} IR triplet lines share a lower level of high excitation 
potential and, hence, are expected to form in the deep photosphere. It 
could therefore be expected that 1D-NLTE spectral line synthesis of such 
lines is influenced by the choice of the mixing-length parameter, \mlp, 
which quantifies the convective efficiency and thus controls the temperature
structure of the lower photosphere.  To probe the possible influence 
of \mlp, we  computed a grid of 1D-NLTE line profiles using solar \LHD\ 
model atmospheres computed with the different values of \mlp. This grid of
synthetic spectra is used to perform the same line profile fitting as 
described in Sect.\,\ref{sect:proffit}.
   
Despite the fact that the fits are quite bad in general (see discussion in
Sect.\,\ref{sect:proffit} and Fig.~\ref{fig:ctlProf1D_mic12}), we arrived at very similar results regardless of the chosen \mlp.
Table~\ref{table:mlp} shows the results of line profile fitting with 
different values of \mlp\ and fixed $\vmic = 1.2$\,km/s (assuming
$\vmic = 0.8$\,km/s yields similar conclusions). Varying \mlp\ 
between 0.5 and 1.5 changes the best-fitting \logAO\ by at most
0.006\,dex (for the weakest triplet component). Also, the reduced 
\chisq\ values of the best fit are  only marginally influenced. 
We conclude that \logAO\ is very weakly dependent on the choice of \mlp, and
the goodness of the fits cannot be improved by modifying \mlp. Hence our
results derived from LHD models with $\mlp = 1.0$ should be representative for
all theoretical 1D mixing-length models of the solar atmosphere.

\begin{table}[h]
\caption {Best-fitting \logAO\ derived from the two sets of observed spectra,
using 1D \LHD\ model atmospheres with different \mlp\ values between 0.5 
and 1.5, and fixed $\vmic = 1.2$\,km/s.} \label{table:mlp} 
 \centering
\begin{tabular}{|c||c|c|c||c|c|c|}
\hline 
$\lambda$,\,nm   & \multicolumn{3}{c|}{WCLC data set} & \multicolumn{3}{c|}{Pereira's data set}\\
\hline 
         & \multicolumn{6}{c|}{\mlp} \\
\hline 
         & 0.5    & 1.0    & 1.5    & 0.5    & 1.0    & 1.5    \\
\hline
777.2    & 8.581  & 8.581  & 8.583  & 8.579  & 8.580  & 8.582  \\
777.4    & 8.592  & 8.593  & 8.595  & 8.596  & 8.597  & 8.600  \\
777.5    & 8.603  & 8.604  & 8.607  & 8.604  & 8.606  & 8.610  \\
\hline
\end{tabular} 
\end{table}

\section{Supplementary plots: line profile fitting results for 1D model 
         atmospheres}
\label{app:1Dprofiles}
In this appendix, we show the observed center-to limb variation 
of the line profiles of the \ion{O}{i} IR triplet, separately for
each component, in comparison with the best-fit theoretical non-LTE 
line profiles derived from various 1D model atmospheres.
The best-fit values of the oxygen abundance, \logAO, are given in the 
legend of each plot, as well as in Table\,\ref{table:profit} (together 
with the best-fit \SH\ values). The best fit correspond to the global minimum 
of \chisq\ as defined by Eq.\,(\ref{eq:chi2profit2}), i.e., simultaneously fitting the 
line profiles at all $\mu$-angles. The quality of the 
fit can be judged from the reduced \chisq\ values provided in the Table, 
as well as from the plots of the intensity difference 
$I^{\rm obs}-I^{\rm calc}$ displayed below the plot of the line profiles.

\begin{figure*}[!ht]
 \centering
 \mbox{\includegraphics[width=17cm,clip=true, trim=0 0mm 0 0mm]
 {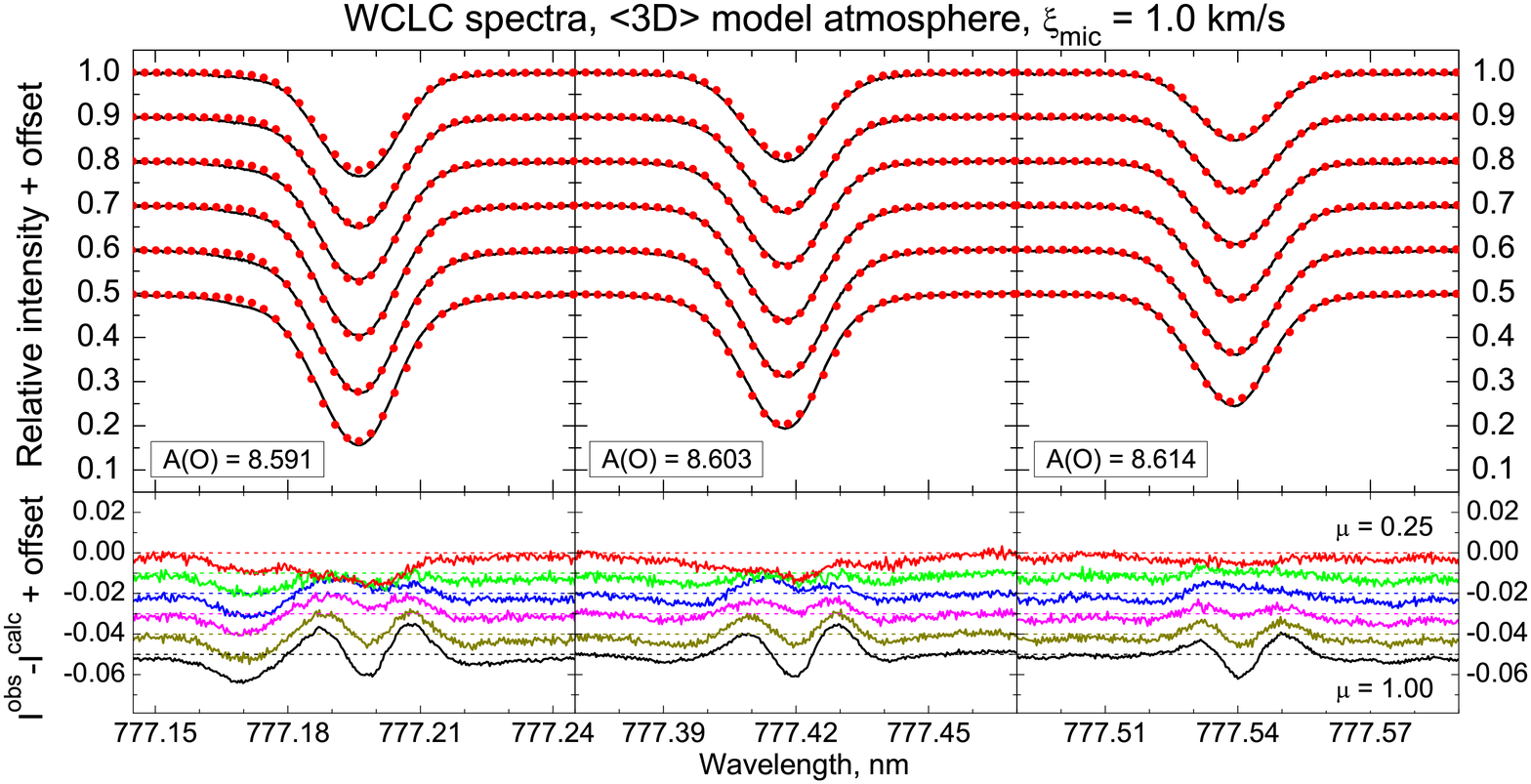}}\\[2mm]
 \mbox{\includegraphics[width=17cm,clip=true, trim=0 0mm 0 0mm]
 {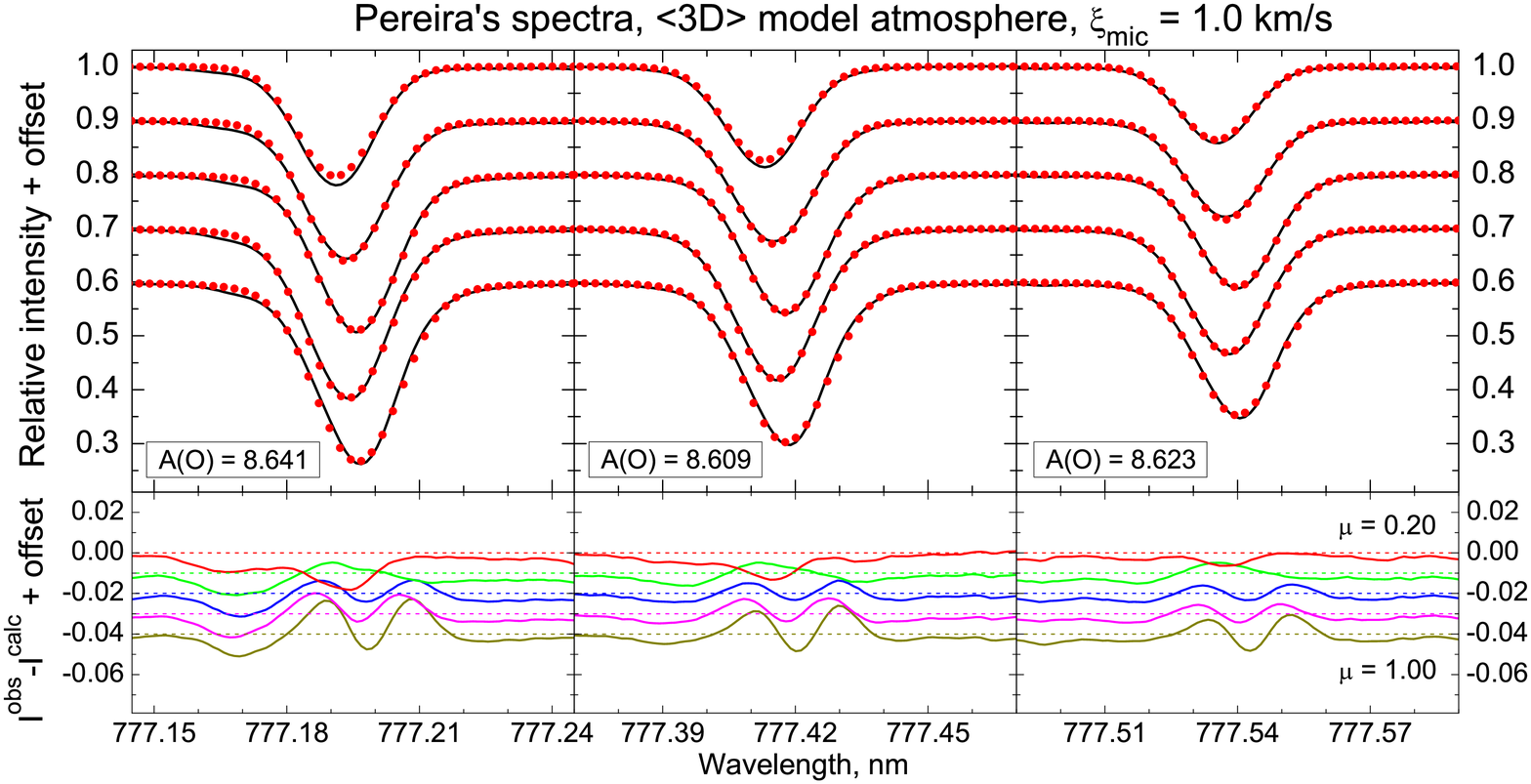}}
 \caption{Same as in Fig.~\ref{fig:ctlProf3D}, except for the averaged 3D
          model atmosphere with $\vmic = 1.0$\,km/s.}
 \label{fig:ctlProf3Dmean_mic10}
\end{figure*}

\begin{figure*}[htb]
 \centering
 \mbox{\includegraphics[width=17cm,clip=true, trim=0 0mm 0 0mm]
 {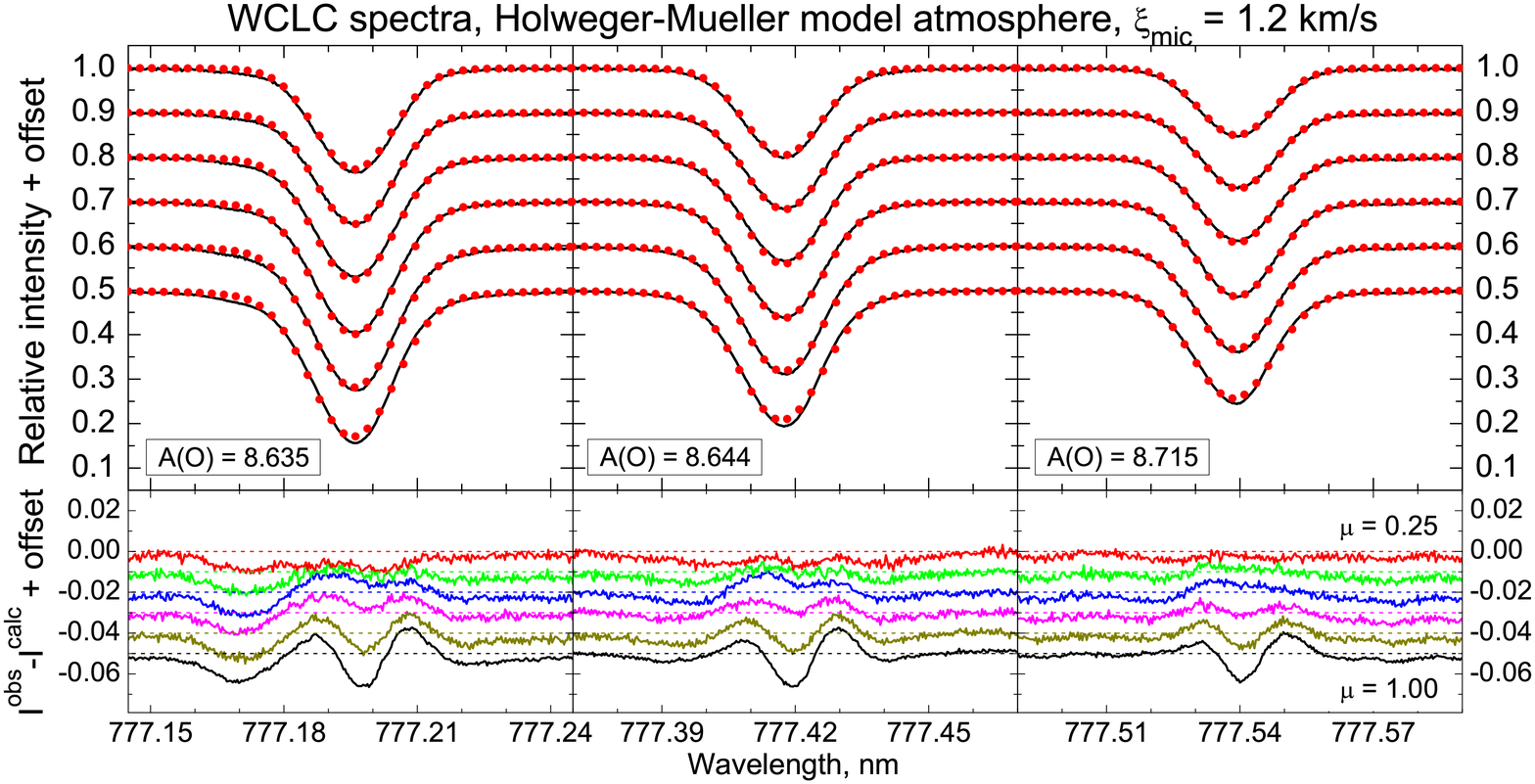}}\\[2mm]
 \mbox{\includegraphics[width=17cm,clip=true, trim=0 0mm 0 0mm]
 {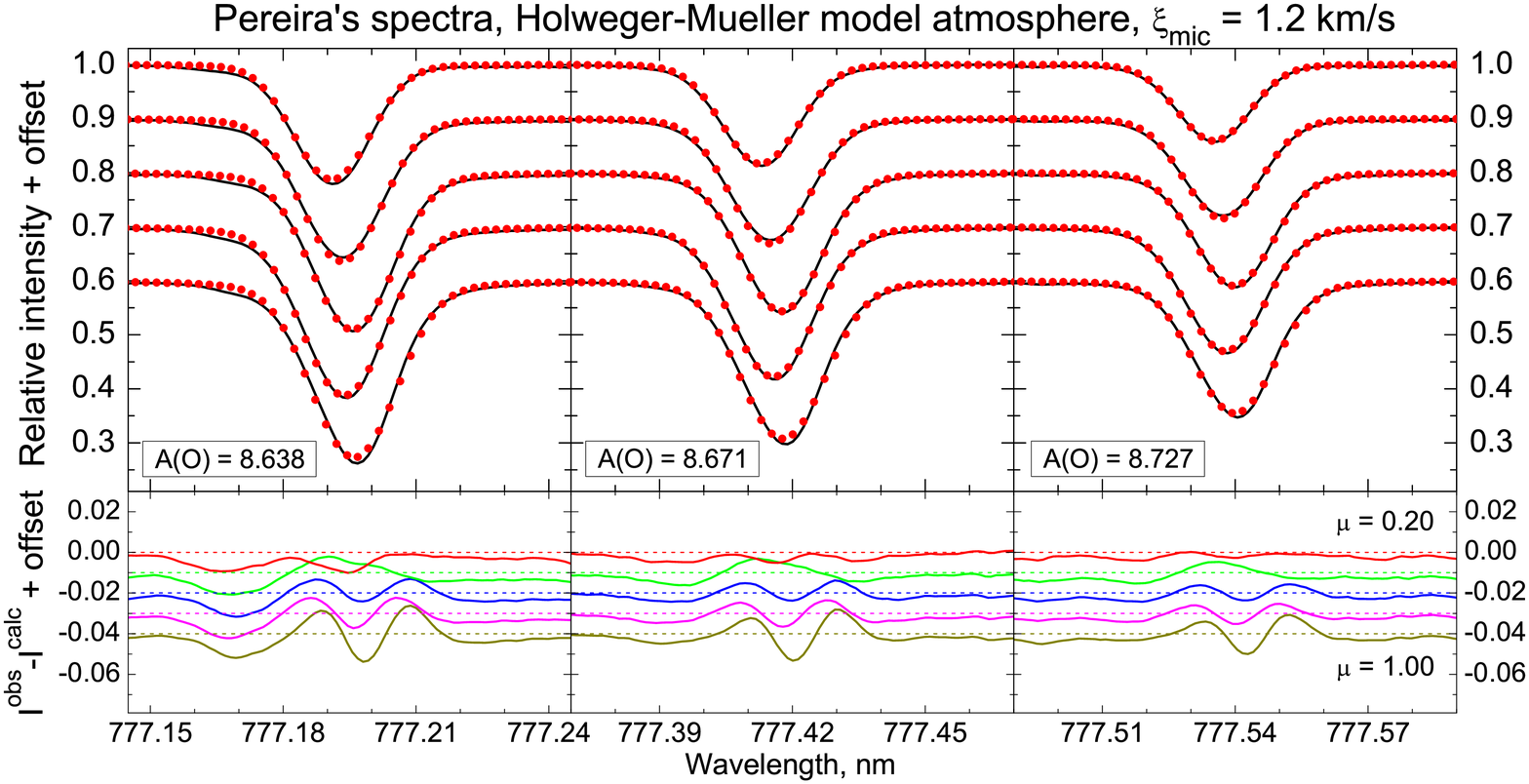}}
 \caption{Same as in Fig.~\ref{fig:ctlProf3D}, except for the Holweger-M\"uller 
          model atmosphere with $\vmic = 1.2$\,km/s.}
 \label{fig:ctlProfHM_mic12}
\end{figure*}

\begin{figure*}[htb]
 \centering
 \mbox{\includegraphics[width=17cm,clip=true, trim=0 00mm 0 0mm]
 {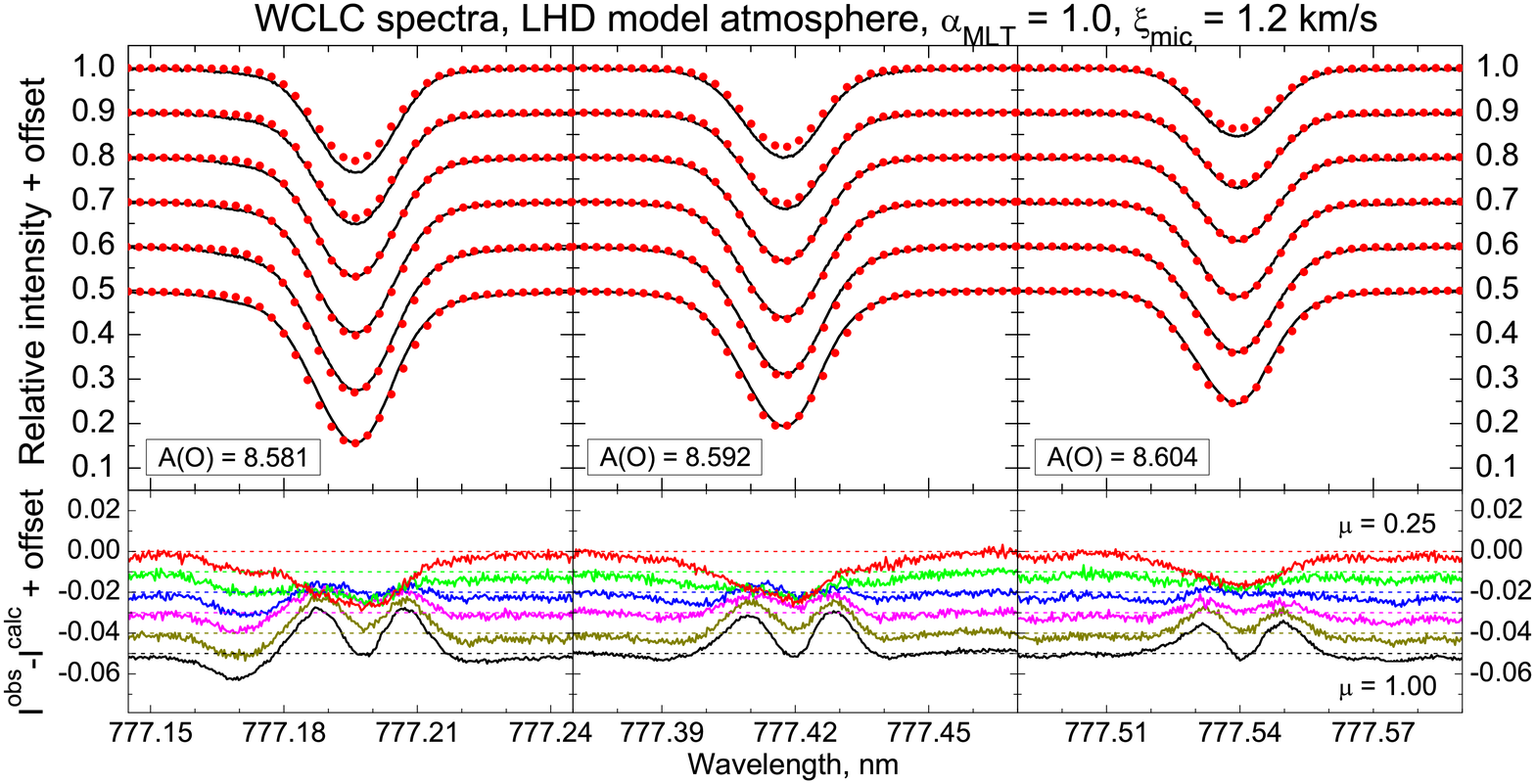}}\\[2mm]
 \mbox{\includegraphics[width=17cm,clip=true, trim=0 00mm 0 0mm]
 {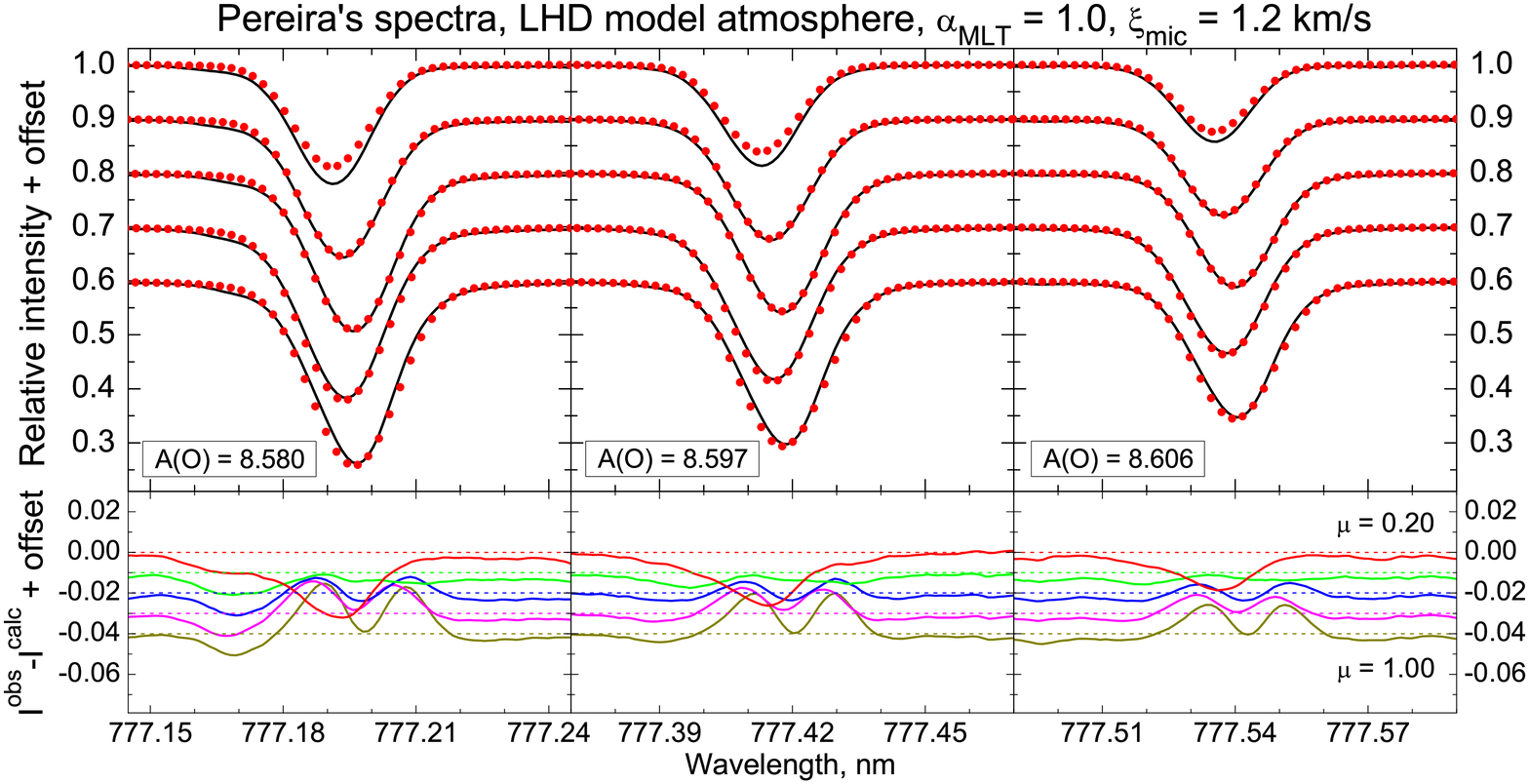}}
 \caption{Same as in Fig.~\ref{fig:ctlProf3D}, except for the \LHD\ model 
          atmosphere with $\mlp = 1.0$ and $\vmic = 1.2$\,km/s.}
 \label{fig:ctlProf1D_mic12}
\end{figure*}


\section{Supplementary plots: equivalent width fitting results}
\label{app:1DEWs}

In this appendix, we show the center-to-limb variation 
of the observed equivalent widths, separately for each component 
of the \ion{O}{i} IR triplet, in comparison with the
theoretical results derived from the 3D and various 1D model atmospheres.
For given \SH=0 and 8/3, the oxygen abundance is computed as the 
\logAO\ value that minimizes the global \chisq\ given by 
Eq.\,(\ref{eq:chi2ewfit}). The best fit is defined as the \SH\ -- \logAO\
combination that corresponds to the global minimum of \chisq. The best-fit
values of \SH\ and \logAO\ are given in Table\,\ref{table:EWfit}. The quality
of the fit can be judged from the reduced \chisq\ values provided in the 
Table, as well as from the plots of the abundance difference 
\mbox{A$(\mu)$ - $\overline{{\rm A}}$}
shown in the bottom panels of each Figure. 
Here A$(\mu)$ is the value of \logAO\ obtained from fitting the line profile
separately at the individual $\mu$-angle, while $\overline{{\rm A}}$ denotes
the \logAO\ value derived from simultaneously fitting the line profiles at all $\mu$-angles.

\begin{figure*}[htb]
 \centering
 \mbox{\includegraphics[width=8.5cm,clip=true]{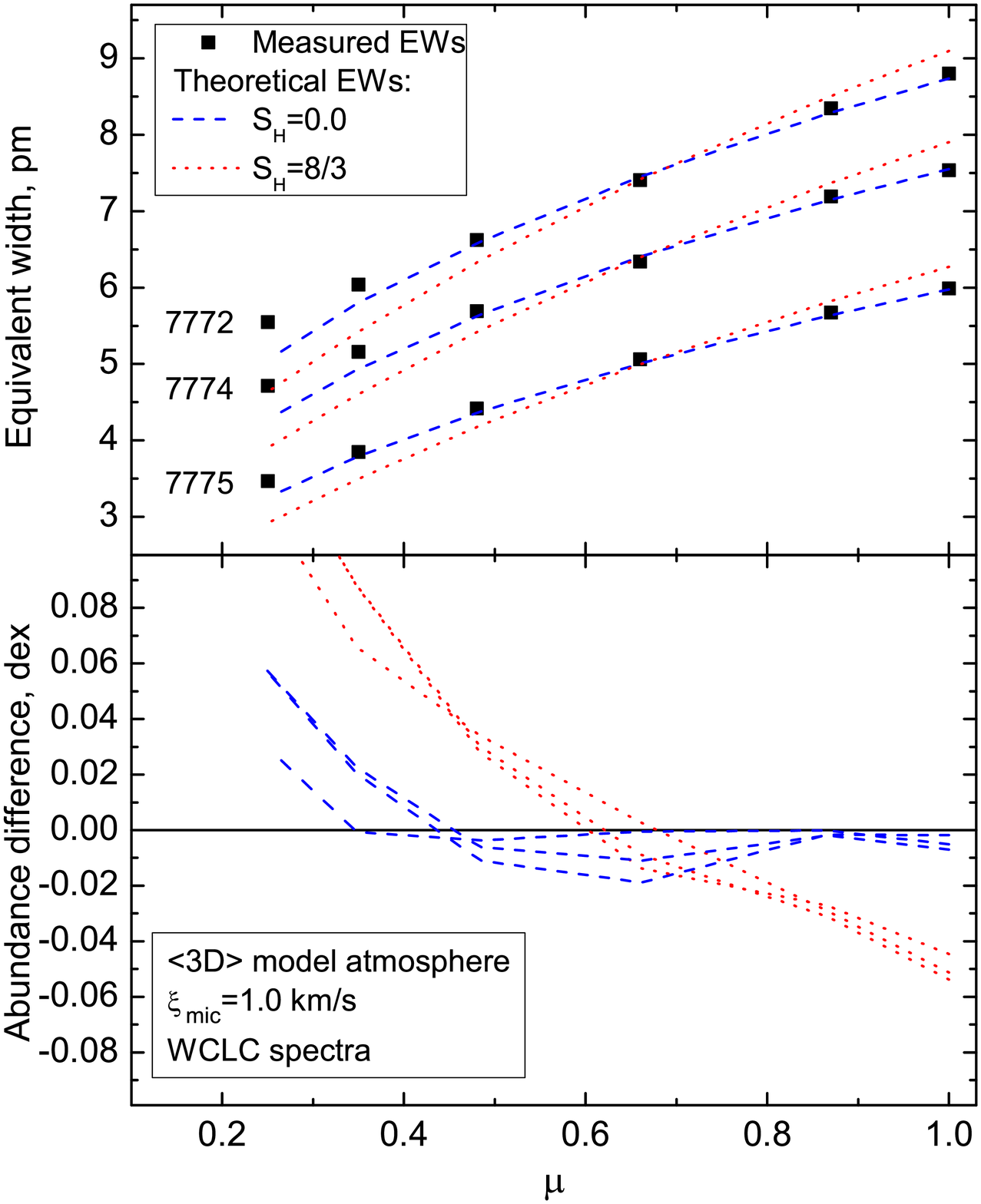}}
 \mbox{\includegraphics[width=8.5cm,clip=true]{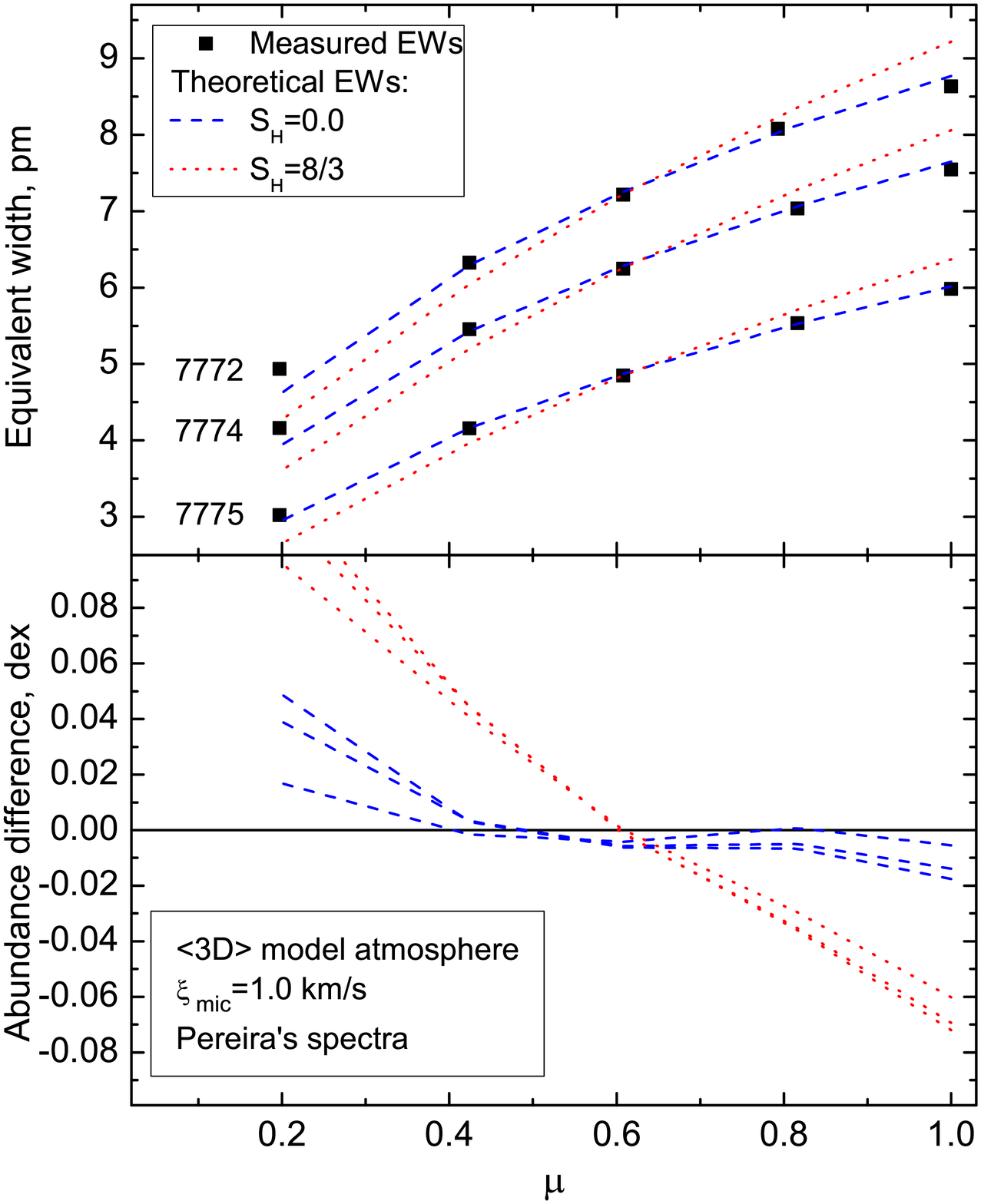}}
 \caption{Same as in Fig.~\ref{fig:ctlEWs}, except for the averaged 3D model 
          atmosphere with $\vmic = 1.0$\,km/s. Best-fit solutions are not 
          shown here, since they would require negative \SH.}
 \label{fig:ctlEWs_3Dmeanmic10}
\end{figure*}

\begin{figure*}[htb]
 \centering
 \mbox{\includegraphics[width=8.5cm,clip=true]{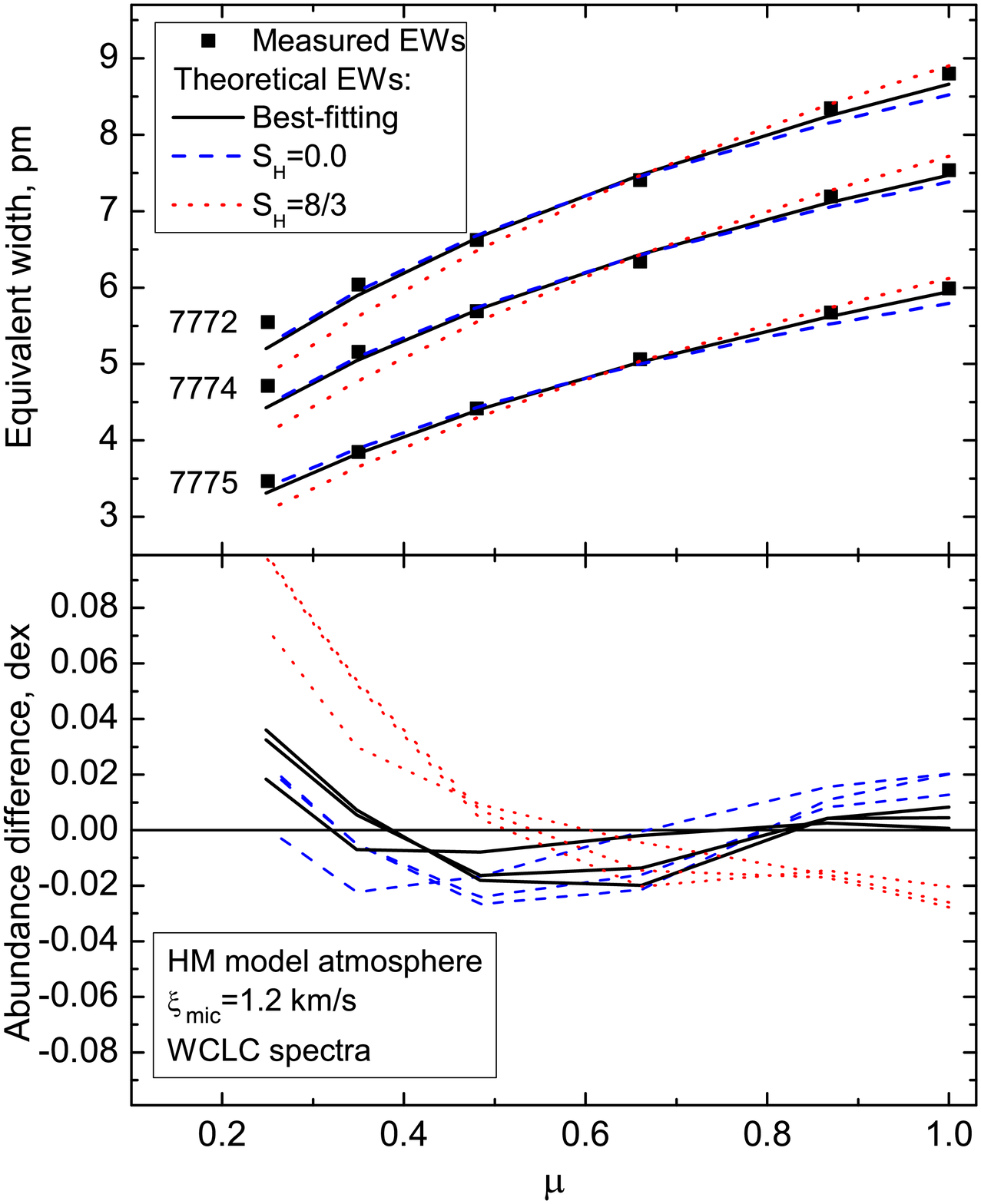}}
 \mbox{\includegraphics[width=8.5cm,clip=true]{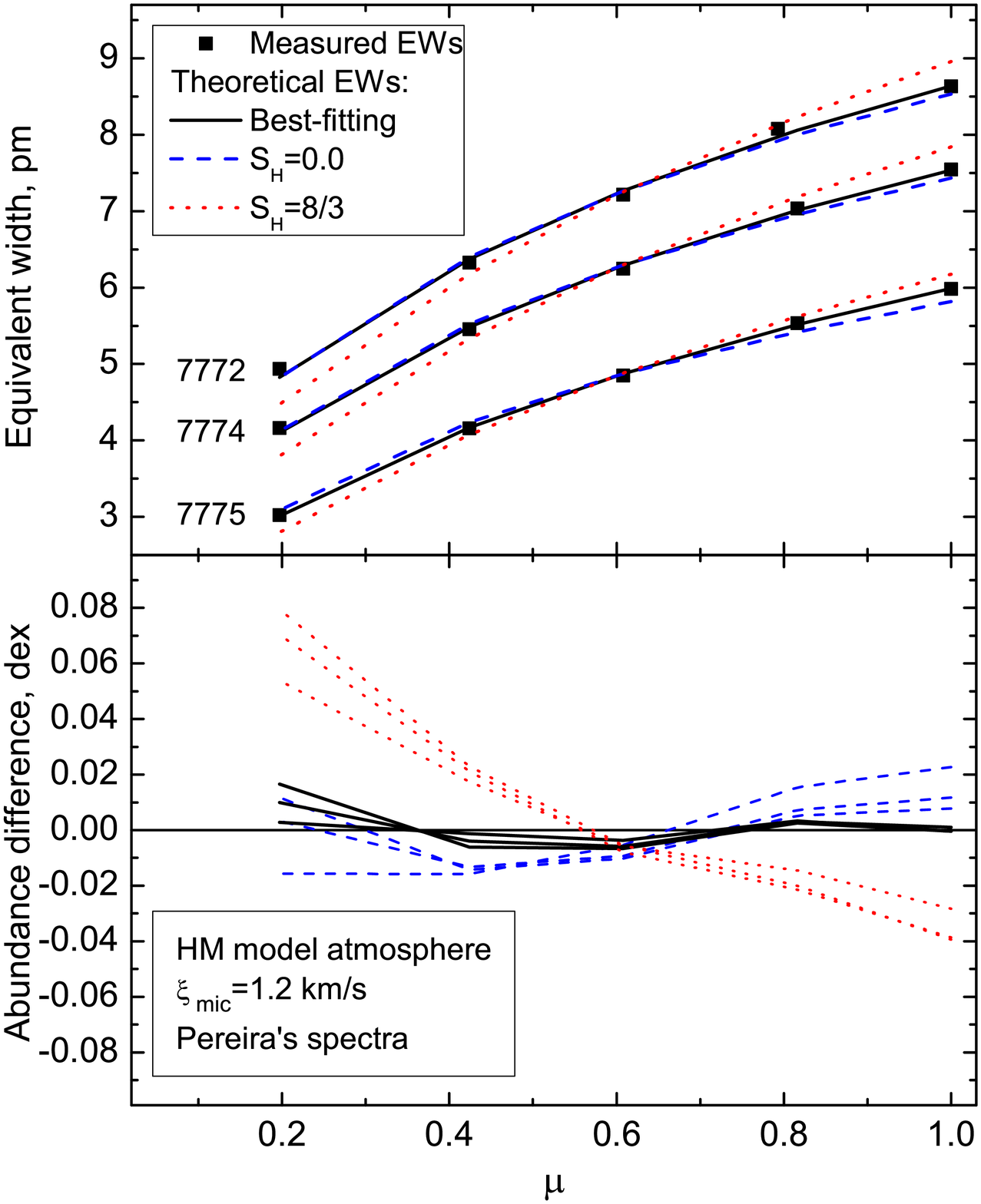}}
 \caption{Same as in Fig.~\ref{fig:ctlEWs}, except for the Holweger-M\"uller 
          model atmosphere with $\vmic = 1.2$\,km/s.}
 \label{fig:ctlEWs_HMmic12}
\end{figure*}

\begin{figure*}[htb]
 \centering
 \mbox{\includegraphics[width=8.5cm,clip=true]{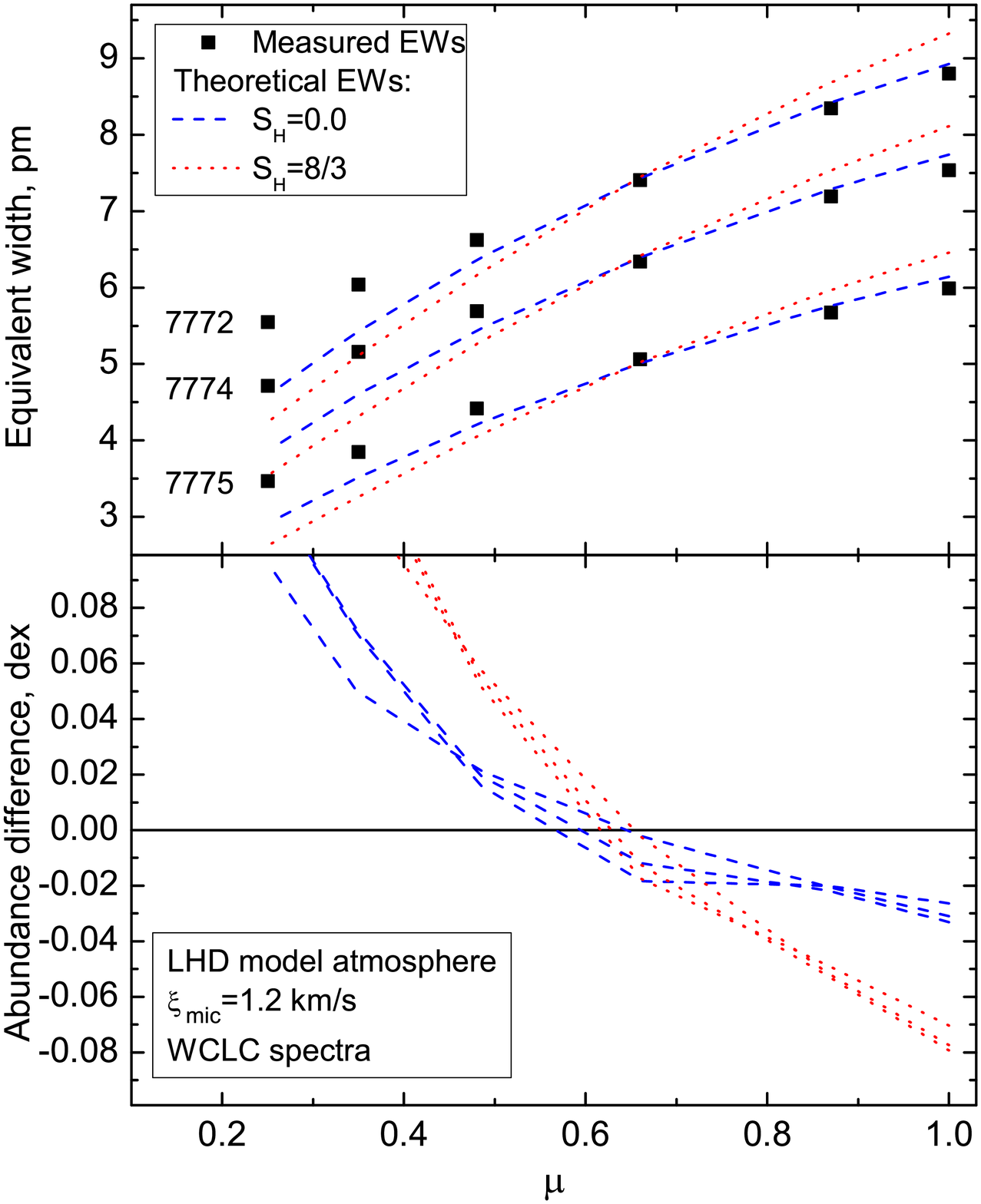}}
 \mbox{\includegraphics[width=8.5cm,clip=true]{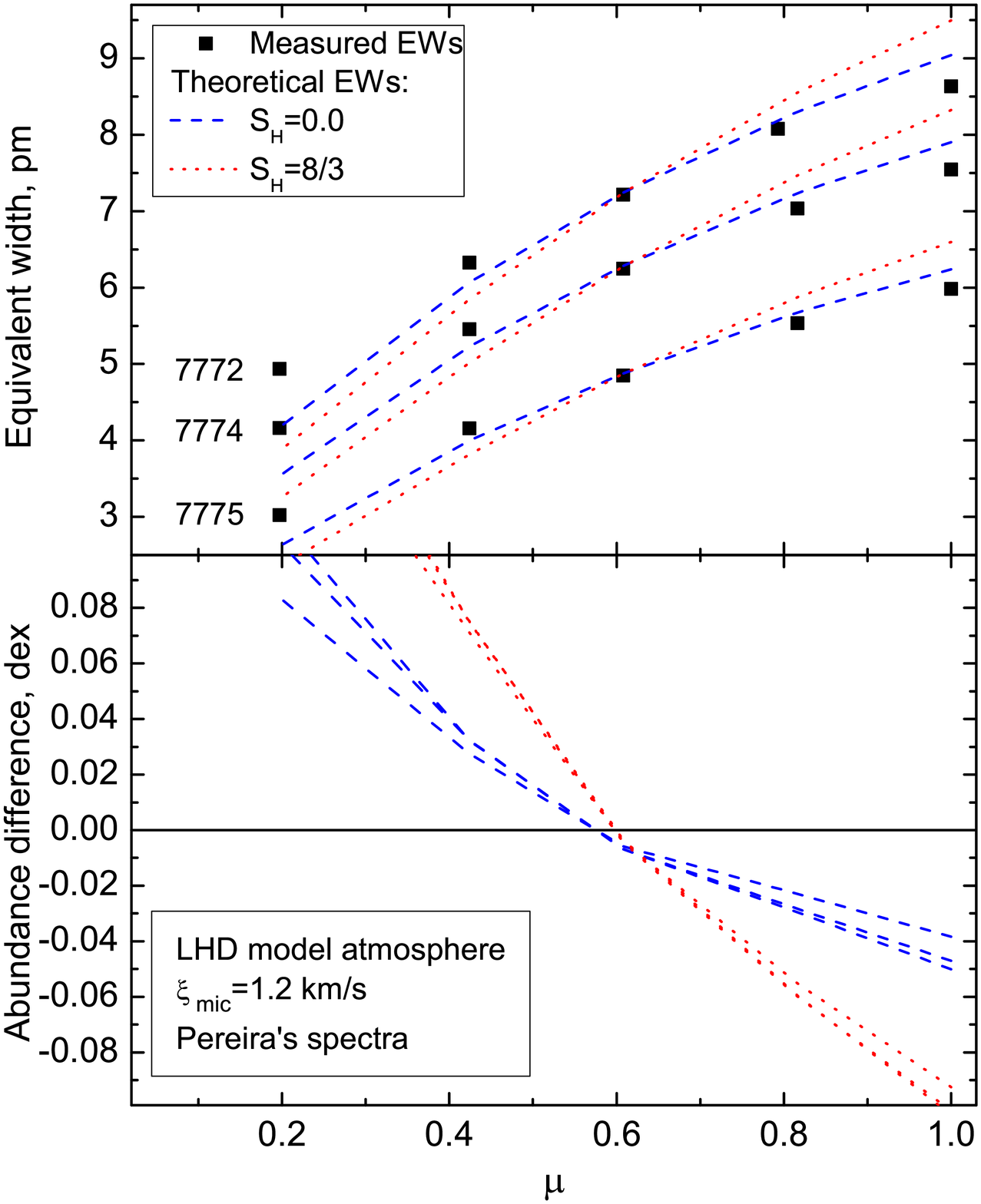}}
 \caption{Same as in Fig.~\ref{fig:ctlEWs}, except for the \LHD\ model 
          atmosphere with $\mlp = 1.0$ and $\vmic = 1.2$\,km/s. Best-fit
          solutions are not shown here, since they would require negative \SH.}
 \label{fig:ctlEWs_LHDmic12}
\end{figure*}

\end{appendix}
   
\end{document}